\documentclass[referee]{aa} %
\usepackage{graphicx}
\usepackage{natbib}
\usepackage{ulem}
\usepackage{longtable}
\usepackage{subfig}
\usepackage{float}
\usepackage[T1]{fontenc}
\usepackage[usenames,dvipsnames]{xcolor}
\usepackage{txfonts}
\newcommand{\teff}{T_\mathrm{eff}}
\begin{document}
    \title{High-resolution spectroscopic atlas of M subdwarfs. Effective temperature and metallicity \footnote{Based on observations made with the ESO Very Large Telescope at the
Paranal Observatory under programme 087.D-0586A and 087.D-0586B} $^{\,,}$\footnote{Tables 1 and 2 along with reduced spectra used for the model comparison are available
at the CDS via anonymous ftp to \\
           cdsarc.u-strasbg.fr (130.79.128.5) or via \\
http://cdsarc.u-strasbg.fr/viz-bin/qcat?J/A+A/545/A85}}
\titlerunning{Effective temperature and metallicity of M subdwarfs.}
\authorrunning{Rajpurohit et al.}

\author{Rajpurohit, A. S.\inst{1}, Reyl\'e, C.\inst{1}, Allard, F.\inst{2}, Scholz, R.-D.\inst{3}, Homeier D.\inst{2}, Schultheis, M.\inst{4}, Bayo, A.\inst{5,}\inst{6}}

\institute{Institut UTINAM, CNRS UMR 6213, Observatoire des Sciences de l'Univers THETA Franche-Comt\'{e}-Bourgogne, Universit\'e de Franche Comt\'{e}, 
Observatoire de Besan\c{c}on, BP 1615, 25010 Besan\c{c}on Cedex, France
\email{arvind@obs-besancon.fr}
\and
Centre de Recherche Astrophysique de Lyon, CNRS UMR 5574, Universit\'{e} de Lyon, \'{E}cole Normale Sup\'{e}rieure de Lyon, 46 all\'{e}e d'Italie,
69364 Lyon Cedex 07, France
\and
Leibniz-Institut f\"ur Astrophysik Potsdam (AIP), An der Sternwarte 16, D-14482 Potsdam, Germany
\and
Universit\'{e} de Nice Sophia-Antipolis, Observatoire de C\^{o}te d'Azur, Lagrange, CNRS UMR 7293, 06304 Nice Cedex 4, France
\and
European Southern Observatory, Alonso de C\'ordova 3107, Vitacura, Santiago, Chile
\and
Max Planck Institut fur Astronomie, Konigstuhl 17, 69117, Heidelberg, Germany
   }

   \date{Received ...; accepted ...}

% \abstract{}{}{}{}{} 
% 5 {} token are mandatory
 
 \abstract
{M subdwarfs are metal-poor and cool stars. They are important probes of the old galactic populations. However, they remain elusive because of their low luminosity. Observational and modelling efforts are required to fully understand their  physics and to investigate the effects of metallicity in their cool atmospheres.}
{We performed a detailed study of a sample of subdwarfs to determine their stellar parameters and constrain the state-of-the art atmospheric models.}
{We present UVES/VLT high-resolution spectra of three late-K subdwarfs and 18 M subdwarfs. Our atlas covers the optical region from 6400 $\AA$ to
the near-infrared at 8900 $\AA$. We show spectral details of cool atmospheres at very high-resolution (R$\sim$ 40 000) and compare them with synthetic
spectra computed from the recent BT-Settl atmosphere models. }
{Our comparison shows that molecular features (TiO, VO, CaH), and atomic features (Fe I, Ti I, Na I, K I) are well-fitted by current models. We produce a relation of effective temperature versus spectral type over the entire subdwarf spectral sequence.
The high-resolution of our spectra, unable us to perform a detailed comparison of the line profiles of individual elements such as Fe I, Ca II, and Ti I, and we determined accurate metallicities of these stars. These determinations in turn enable us to calibrate the relation between metallicity and molecular-band strength indices from low-resolution spectra.}
{The new generation of models is able to reproduce various spectral features of M subdwarfs. These high-resolution spectra allowed us to separate the atmospheric parameters (effective temperature, gravity, metallicity), which is not possible when using low-resolution spectroscopy or photometry.}


   \maketitle
%
%________________________________________________________________

\section{Introduction}
M subdwarfs are metal-poor, low-luminosity dwarfs. Very few are known, and as a result the bottom-end of the main sequence is very poorly constrained for metal-poor stars. The search for M subdwarfs is hampered not only by the fact that metal-poor stars are rare and intrinsically faint, but also because late-type M subdwarfs do not show exceptionally red colours, as ultracool-M dwarfs and brown dwarfs do \citep{Lepine2003a}. %They appear less luminous because their atmosphere is deficient in metals \citep{Baraffe1997}. 
%Due to their lower metallicity and intrinsic faintness, M subdwarfs lie below their solar-metallicity counterpart in the Hertzsprung-Russel diagram. 
Subdwarfs are typically very old (10 Gyr or older) and belong to the old Galactic populations: old disc, thick disc and spheroid, as shown by their spectroscopic features, kinematic properties and ages \citep{Digby2003,Lepine2003a, Burgasser2003}. Because the low-mass subdwarfs, with their extremely long nuclear-burning lifetimes, were presumably formed early in the Galaxy's history, they are important tracers of the Galactic structure and chemical enrichment history. In addition, detailed studies of their complex spectral energy distributions give new insights on the role of metallicity in the opacity structure, chemistry, and evolution of cool atmospheres, and on fundamental questions on spectral classification and temperature-vs-luminosity scales.  \cite{Gizis1997} proposed a first classification of M subdwarfs (sdM) and extreme subdwarfs (esdM) based on TiO and CaH band strengths in low-resolution optical spectra. \cite{Lepine2007} have revised the adopted classification, and proposed a new classification for those most metal-poor, the ultra-subdwarfs (usdM).  \cite{Jao2008} compared model grids with the optical spectra to characterize the M subdwarfs by three parameters: temperature, gravity, and metallicity, and thus gave an alternative classification scheme of subdwarfs. 

Spectroscopic studies of sdM  at high-resolution have proven to be a difficult task. In the low-temperature regime occupied by these stars, the optical spectrum is covered by a forest of molecular lines, hiding or blending most of the atomic lines used in spectral analysis. However, over the past decade, stellar atmosphere models of very low mass stars have made much progress by exploring metallicity effects \citep{Allard2012,Allard2013}. One of the most important recent improvement is the revision of the solar abundances \citep{Asplund2009,Caffau2011}.

Rapid progress in the investigation of cool atmospheres is expected thanks to the advent of eight-meter-class telescopes that allow high-resolution spectroscopy of these faint targets. In high-resolution spectra, access to weak lines allows us to determine the metallicity separately from the other main parameters gravity and effective temperature. The pressure changes equally affect all atmospheric parameters and hence the various absorption bands. The determination of gravity from the pressure-broadened wings is expected to be much more accurate than comparing colour ratios from photometry or low-resolution spectra.  Thus it is necessary to achieve a very good fit in all important absorbers to determine atmospheric properties, because the chemical complexity of these atmospheres reacts sensitively to the main opacities. Descriptions of these stars therefore need validations by comparison with high-resolution spectroscopic observations.

Measuring metallicities for M dwarfs is also challenging. Over the entire M dwarf sequence, the effective temperature ranges from about 4000 K to 2400  K \citep[see e.g.][]{Rajpurohit2013}. With decreasing temperature, the spectra show increasingly abundant diatomic and triatomic molecules. % (TiO, VO, H$_2$O, CO, FeH, CaH and MgH). 
 In particular, the TiO and H$_2$O bands have complex and extensive absorption structures, creating a pseudo-continuum that let pass only the strongest, often resonance, atomic lines. 
%However, recent advances in model atmospheres of low-mass stars have boosted the number of studies deriving accurate metallicities for M dwarfs and subdwarfs. 
\cite{Woolf2005} and \cite{Woolf2009} obtained metallicities from high-resolution spectra by measuring equivalent widths of atomic lines in regions less dominated by molecular bands.
Metallicities have also been 
obtained in binaries containing an M-type and a solar-type star from the much better understood spectra of the latter
%. Assuming they are sharing a common metallicity that reflects the composition of the molecular cloud in which they formed, 
%where the metallicity is measured on the much better understood spectrum of the solar-type star 
\citep{Bonfils2005,Bean2006a,Bean2006b}. These results, together with the abundances obtained from high-resolution spectra, are combined to calibrate metallicities using photometry or molecular indices \citep{Bonfils2005,Woolf2009,Casagrande2008}.

Metal-poor stars are rare in the solar neighbourhood, and the current sample of local subdwarfs is very limited.  High-resolution spectra of M dwarfs were shown by \cite{Tinney1998}  on the full optical range, but such observations are not available for the M subdwarfs on the whole temperature sequence.
In this paper we present the first high-resolution optical atlas of stars that covers the whole sdM, esdM, and usdM sequence. It consists of 2 sdK, 11 sdM, 1 esdK, 5esdM, and 2 usdM observed with UVES at VLT. Using the most recent \texttt{PHOENIX} BT-Settl stellar atmosphere models, we performed a detailed comparison with our observed spectra. In this study we compare the models with the high-resolution spectra of sdM at subsolar metallicities and assign effective temperatures. We derive metallicities based on the best fit of synthetic spectra to the observed spectra and compare in detail the line profile of Fe I, Ca II, Ti I, and Na.
 
High-resolution spectroscopic observations are described in Section \ref{atlas} and the model atmospheres used for comparison are presented in Section \ref{S_mod}. The comparison between observed and synthetic spectra and the derived stellar parameters are given in Section \ref{smd}. In Section \ref{disc}, we present effective temperature versus colours and spectral-type relations, and the metallicity calibrations. The conclusion is given in Section \ref{ccl}.  

\section{High-resolution spectral atlas of M subdwarfs}
\label{atlas}
High-resolution spectroscopy is a very important tool for understanding the physics of stars. Identifying of atomic and molecular absorption features in the spectra of M subdwarfs is important they can be used to constrain the main stellar parameters $\teff$, $\mathrm{log}\,g$, [Fe/H]. The determination of gravity from its effect on the pressure-dependent wings of the saturated atomic absorption lines (predominantly of the alkali elements) is expected to be much more accurate than comparing colour ratios from low-resolution spectra \citep{Reiners2007a}. 

We present a high-resolution spectral atlas of 21 very low mass objects that we selected to cover the entire M subdwarf spectral range, including 6 extreme and 2 ultra subdwarfs. The name, spectral type, and near-infrared photometry of these objects are given in Table~1. The photometry was taken from the Two Micron All Sky Survey \citep[2MASS,][]{Skrutskie2006}. The spectral types and spectral indices were found in the literature.

\begin{table*}
\centering
\caption{Spectral types, near-infrared photometry, and spectral indices of our sample. The photometry and coordinates are taken from 2MASS, at epochs between 1997 and 2001. References are given for spectral types first and spectral indices next.}
\begin{tabular}{lcccccccccccc}
\hline
Name& $\alpha$ & $\delta$ &SpT & J&      H&      K$_s$& 	TiO5 &CaH2	&CaH3 &Ref.\\
\hline

LHS 72    &23 43 13.5&    $-$24 09 50    &sdK4&  9.61&     9.04&      8.82    &-- &-- &-- &a,-- \\
LHS 73    &23 43 16.6& $-$24 11 14      & sdK7&              10.11&     9.59&         9.37&     0.960 &0.847 &0.901 &b,b\\
G 18-37&    22 14 55.5&  +05 42 37&    esdK7&  12.72&    12.20&     12.02    &-- &-- &-- &c,-- \\%SDSS2214
APMPM J2126-4454&21 26 23.9&$-$44 53 34&    sdM0&  12.65&    12.11&     11.92&-- &-- &-- &d,-- \\
LHS 300&    11 11 13.8&$-$41 05 33&        sdM0&      10.48&    10.01&     9.80    &    1.031&     0.886&   0.943 &e,e\\
LHS 401&    15 39 39.1& $-$55 09 10&sdM0.5&  10.15&    9.60&     9.41& 0.997&   0.911&   0.958& e,e    \\
LHS 158&    02 42 02.8&$-$44 30 59&    sdM1&  10.43&    9.94&     9.73& 0.732&   0.639&   0.829& e,e   \\
LHS 320&    12 02 33.7&+08 25 51&        sdM2&      10.74&10.18& 9.99 & 0.620&    0.512&   0.751&f,f   \\
LHS 406&    15 43 18.4    &$-$20 15 31&    sdM2&      9.78&     9.23&     9.02 & 0.686&   0.576&   0.789&e,e   \\
LHS 161&    02 52 45.6&+01 55 50&        esdM2&      11.71&    11.20&     11.00& 0.889 &0.689 &0.817 &  f,e  \\
LP 771-87 &03 07 34.0&$-$17 36 38&    usdM2&  15.60& 15.29& 14.90   &--		&--	&--		&g,--	 \\%2MASS 0307-1736
LHS 541&    23 17 06.0&$-$13 50 53&    sdM3&              13.03&    12.56&     12.41& 0.730&   0.467&   0.664& h,h    \\
LHS 272&    09 43 46.3&    $-$17 47 07    &sdM3& 9.62&     9.12&          8.87& 0.678 &0.441 &0.669 &f,f    \\
LP 707-15&    01 09 54.1&$-$10 12 13&    esdM3&        12.94&    12.40&   12.16   &-- &-- &-- &c,-- \\%SDSS0109
LSR 1755+1648&17 55 32.8&+16 48 59&        sdM3.5&      11.35&    10.89&     10.63& 0.486&     0.419&   0.664& i,i\\
LHS 375&    14 31 38.3    &$-$25 25 33&    esdM4&      12.15&    11.67&     11.51   & 0.829&     0.372&   0.547& f,f \\
LHS 1032&00 11 00.8&+04 20 25&        usdM4.5&      14.34&    13.81&     13.76& 0.944&     0.374&   0.514&   j,j    \\
SSSPM J0500-5406 &05 00 15.8&$-$54 06 27&    esdM6.5&      14.44&    14.12&     13.97 & 0.755	&0.220	&0.331		&h,h   \\
LHS 377&    14 39 00.3&+18 39 39&    sdM7&  13.19&    12.73&     12.48& 0.232&   0.205&    0.396&   f,f \\
APMPM J0559-2903&05 58 58.9&$-$29 03 27&    esdM7&  14.89&    14.45&     14.46   & 0.600      &0.210&	0.320& l,l\\
SSSPM J1013-1356 &10 13 07.3&$-$13 56 20&    sdM9.5&      14.62&    14.38&     14.40& 0.208&   0.116&   0.200& m,m   \\
\hline
\end{tabular}

(a)  \cite{Rodgers1974,Bidelman1985}
--
(b)  \cite{Reyle2006}
--
(c)  S. L\'epine, private communication
--
(d)  \cite{Scholz2002}
--
(e)  \cite{Jao2008}
--
(f)  \cite{Gizis1997}
--
(g)  \cite{Kirkpatrick2010}
--
(h)  \cite{Dawson2000}
--
(i)  \cite{Lepine2003}
--
(j)  \cite{Lepine2007}
--
(k)  \cite{Burgasser2006}
--
(l) \cite{Schweitzer1999}
--
(m) \cite{Scholz2004}

\end{table*}

%\hline
%\end{tabular}
%\end{table*}

\subsection{Observation and data reduction}
%\subsection{Objects and format of the atlas}

The observations were carried out in visitor mode during April and September 2011 with the optical spectrometer UVES \citep{Dekker2000} on the Very Large Telescope (VLT) at the European Southern Observatory (ESO) in Paranal, Chile.  UVES was operated in dichroic mode using the red arm with non-standard setting centred at 830 nm. This setting covers the wavelength range 6400 $\AA$-9000 $\AA$, which contains various atomic lines such as Fe I, Ti I, K II, Na I and Ca II and is very useful for the spectral synthesis analysis. Between the two CCDs of the red arm, the spectra have a gap from 8200 $\AA$ to 8370 $\AA$. The Na doublet at 8190 $\AA$ lies just blueward of the gap. The spectra were taken with a slit width of 1.0$^{\prime\prime}$, yielding a nominal resolving power of $R = 40 000$. The signal-to-noise ratio varies over the wavelength region according to the object's spectral energy distribution and detector efficiency. It reaches 30 to more than 100, depending on the magnitude of the targets in most of the spectral region so that the dense molecular and atomic absorption features are clearly discernible from noise. Data were reduced using the software called Reflex for UVES data, which runs standard ESO pipelines modules. 

The spectra are shown in Fig. \ref{Fig:1} for a representative sample along the sdM sequence, and in Fig. \ref{Fig:2} for the esdM and usdM targets.
The spectra were not corrected for terrestrial absorption by O$_3$ and H$_2$O. The O$_3$ features are quite regular and straightforward to distinguish from features in the stars themselves. This is not true for the H$_2$O absorption, which is complex and irregular \citep{Tinney1998}. We therefore show in the upper panels the spectrum of the reference star EG 21, where telluric absorptions appear. The main molecular and atomic features expected in the M subdwarfs are labelled in  Fig.\ref{Fig:1} and Fig.~\ref{Fig:2}.

\begin{figure*}[ht!]
\centering
\subfloat{\includegraphics[width=13.5cm,height=10.0cm]{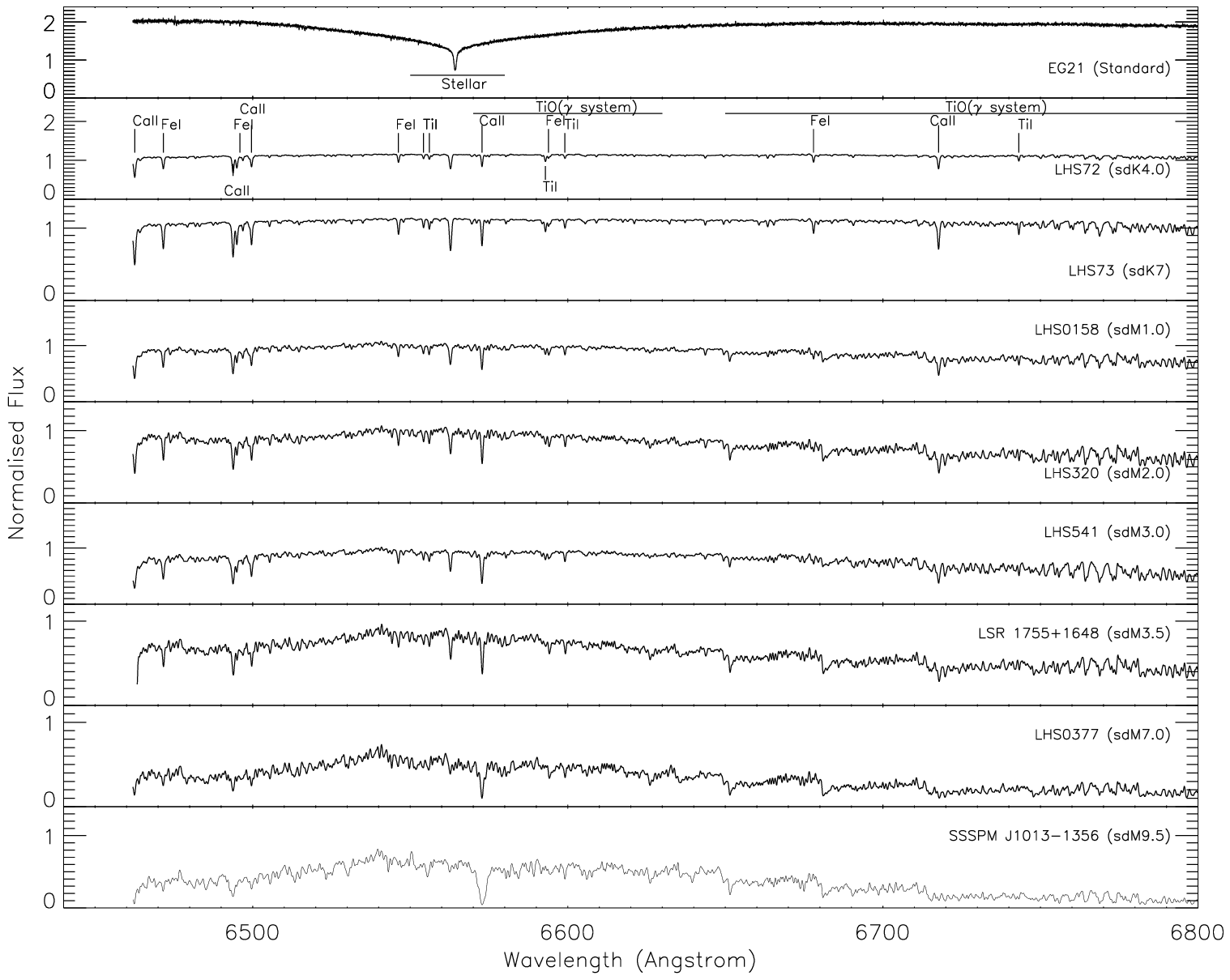}}
\qquad
\subfloat{\includegraphics[width=13.5cm,height=10.0cm]{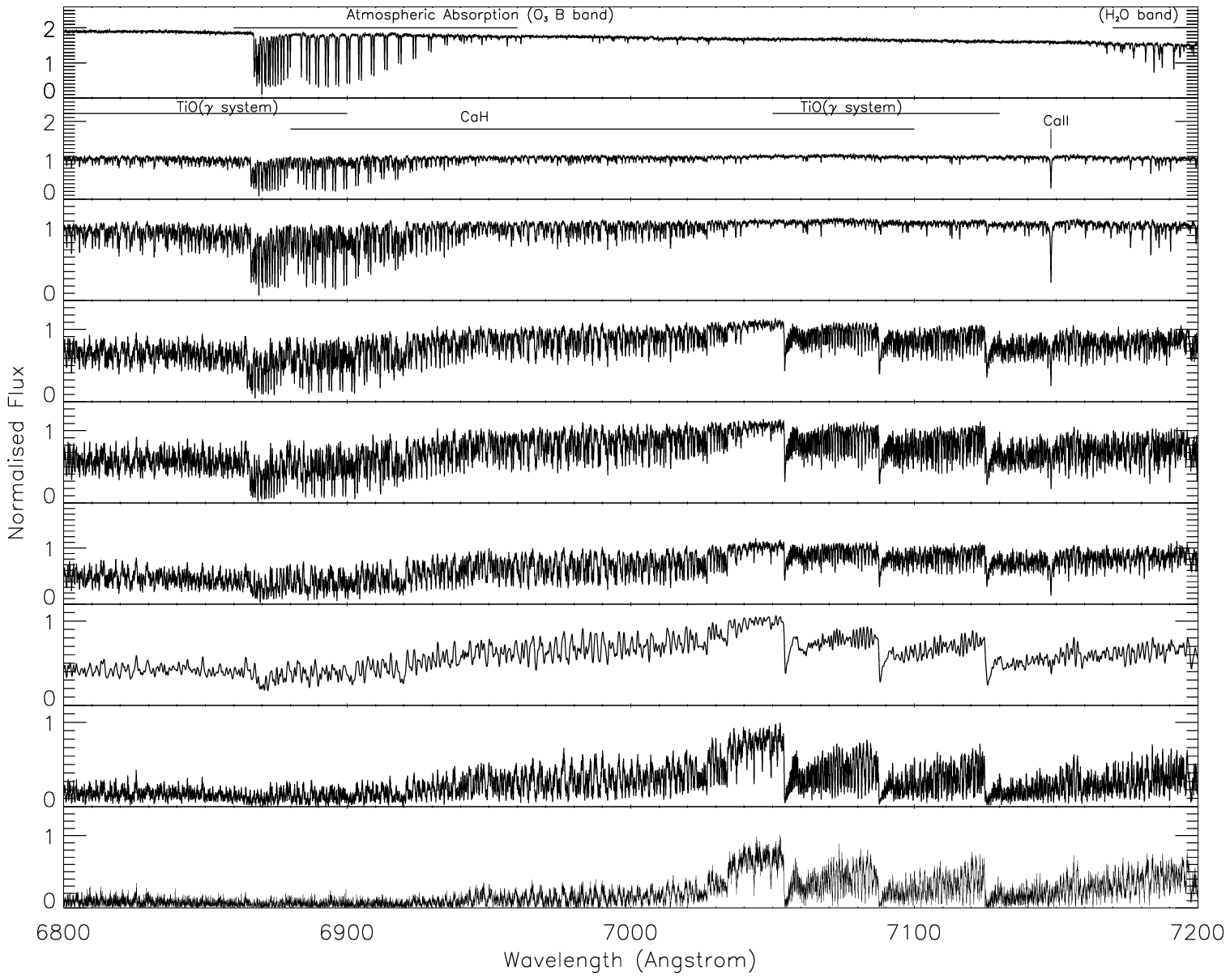}}
\caption{UVES spectra over the sdM spectral sequence.The spectra are scaled to normalize the average flux to unity. The spectrum of the standard star EG 21 shows the location of telluric atmospheric features.}
\label{Fig:1}
\end{figure*}

\begin{figure*}[ht!]
\ContinuedFloat
\centering
\subfloat{\includegraphics[width=13.5cm,height=10.0cm]{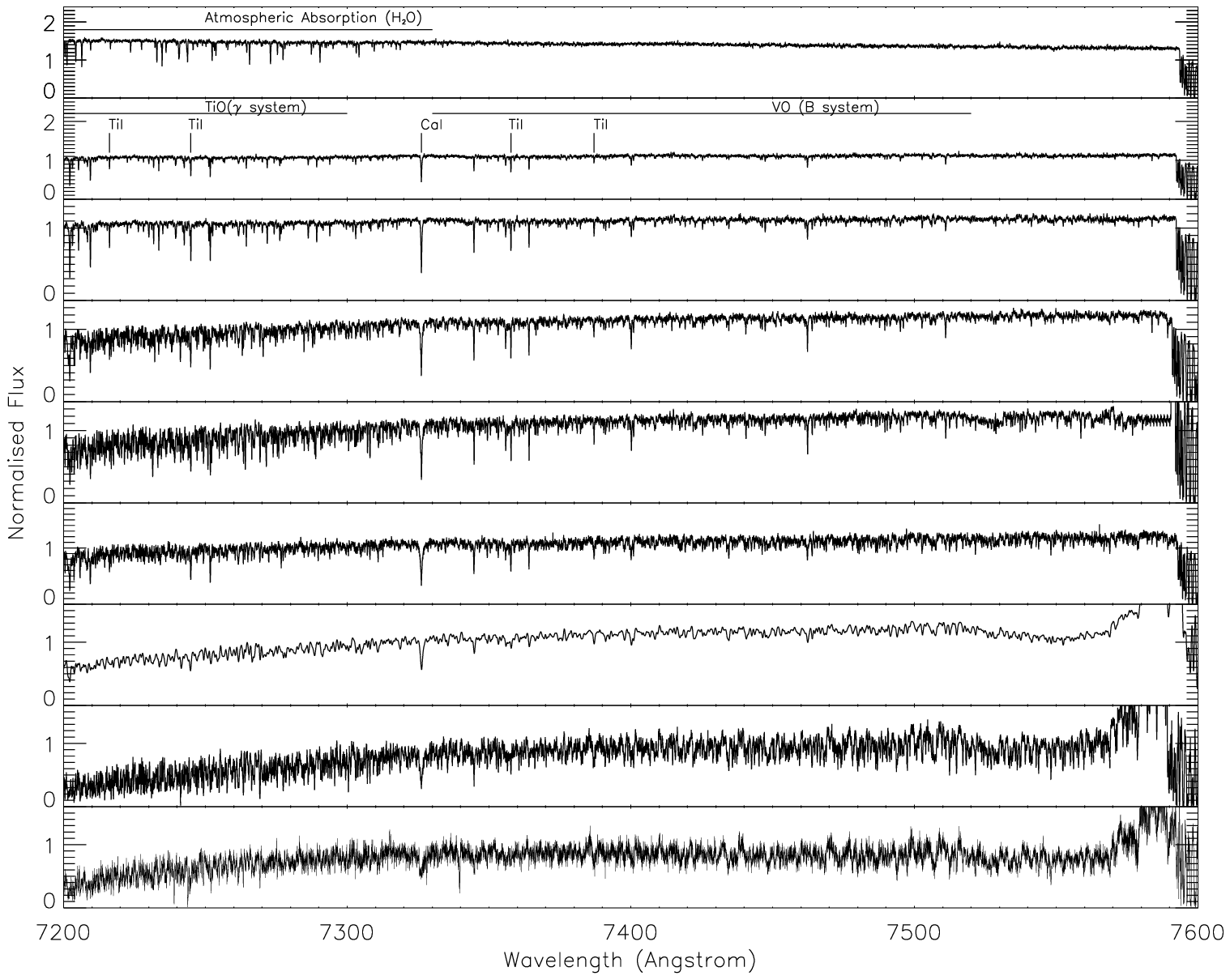}}
\qquad
\subfloat{\includegraphics[width=13.5cm,height=10.0cm]{7600-8230.ps}}
\caption{Continued.}
\label{Fig:1}
\end{figure*}

\begin{figure*}[ht!]
\ContinuedFloat
\centering
\subfloat{\includegraphics[width=13.5cm,height=10.0cm]{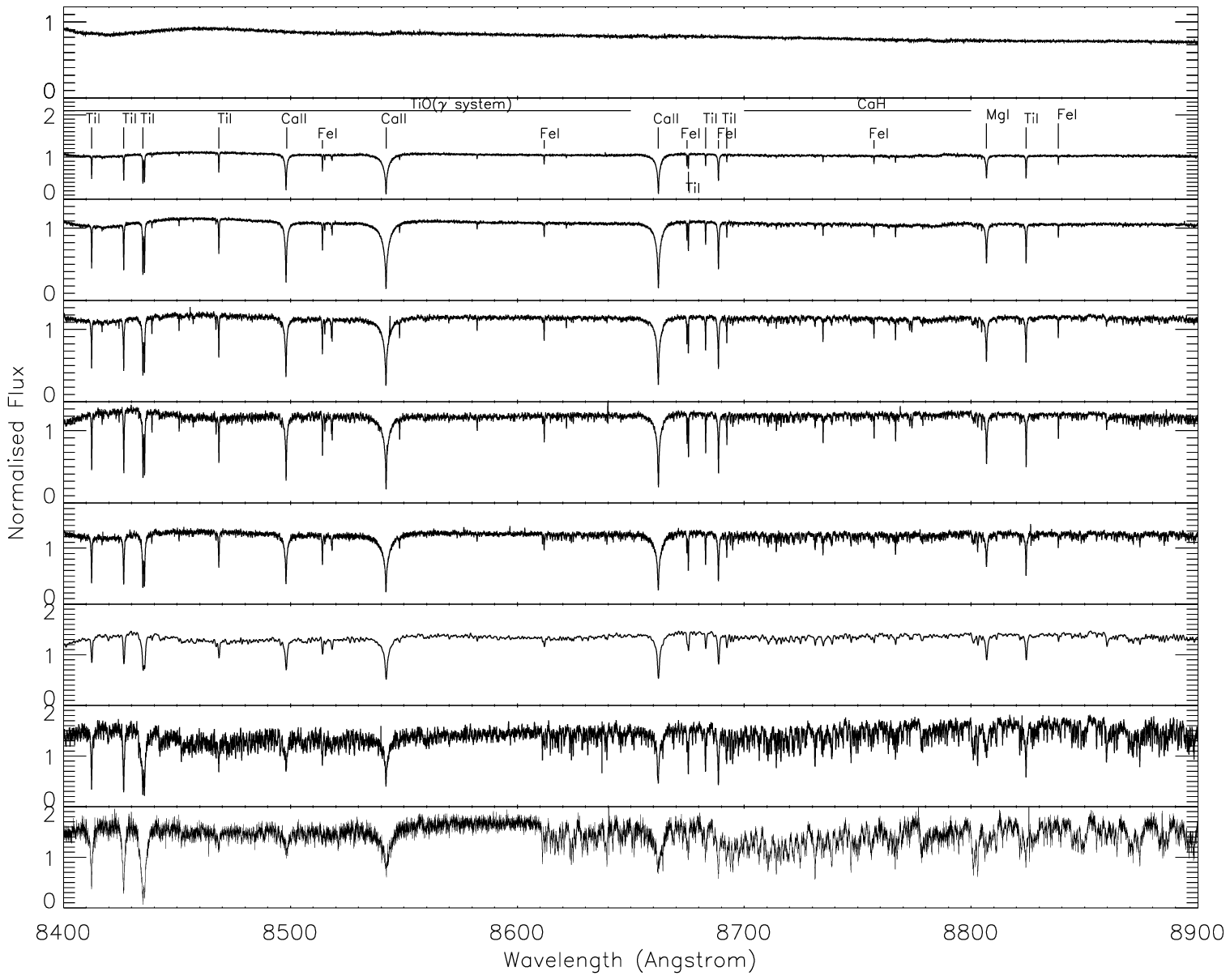}}
\caption{Continued.}
\label{Fig:1}
\end{figure*}

\begin{figure*}[ht!]
\centering
\subfloat{\includegraphics[width=13.5cm,height=10.0cm]{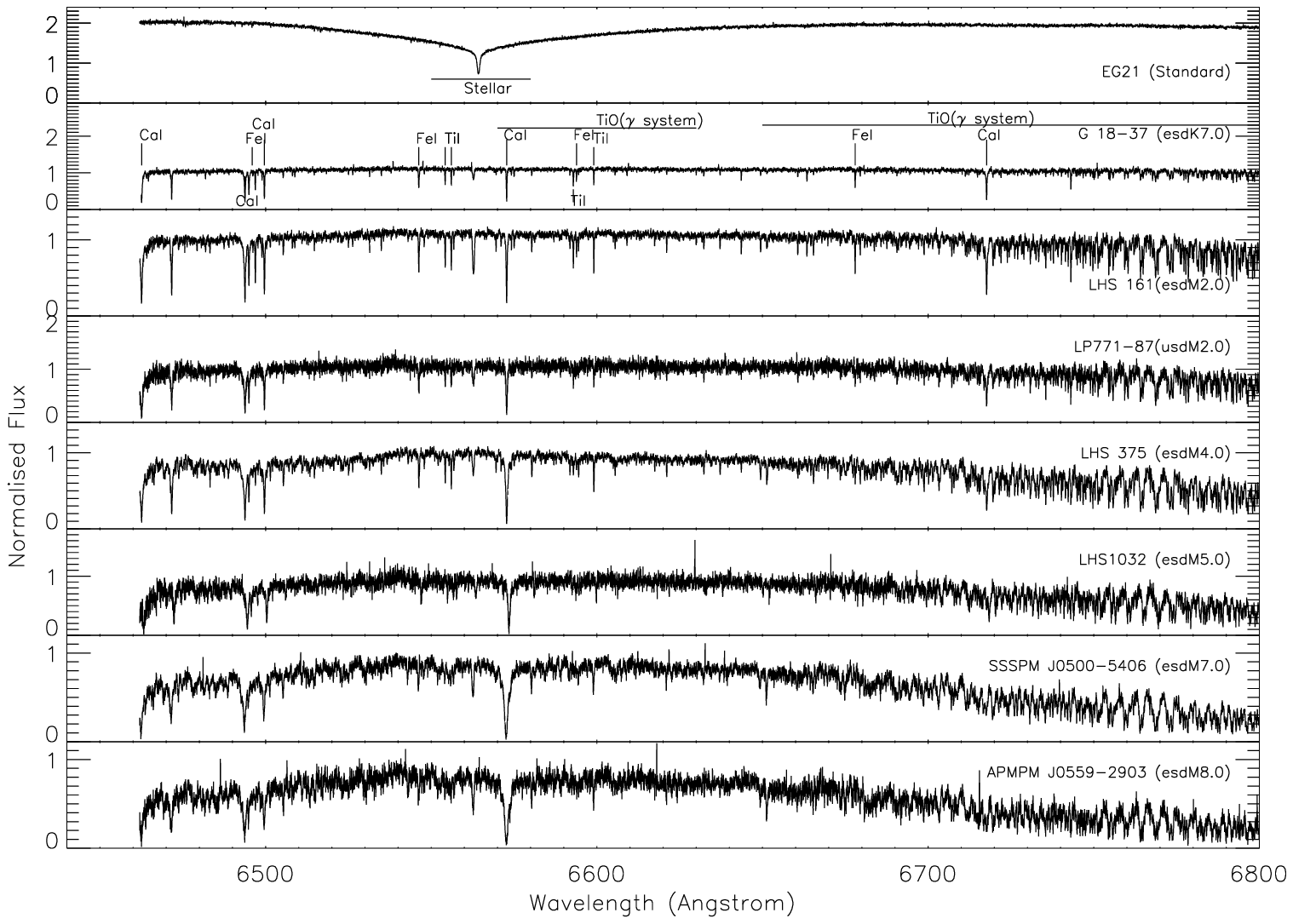}}
\qquad
\subfloat{\includegraphics[width=13.5cm,height=10.0cm]{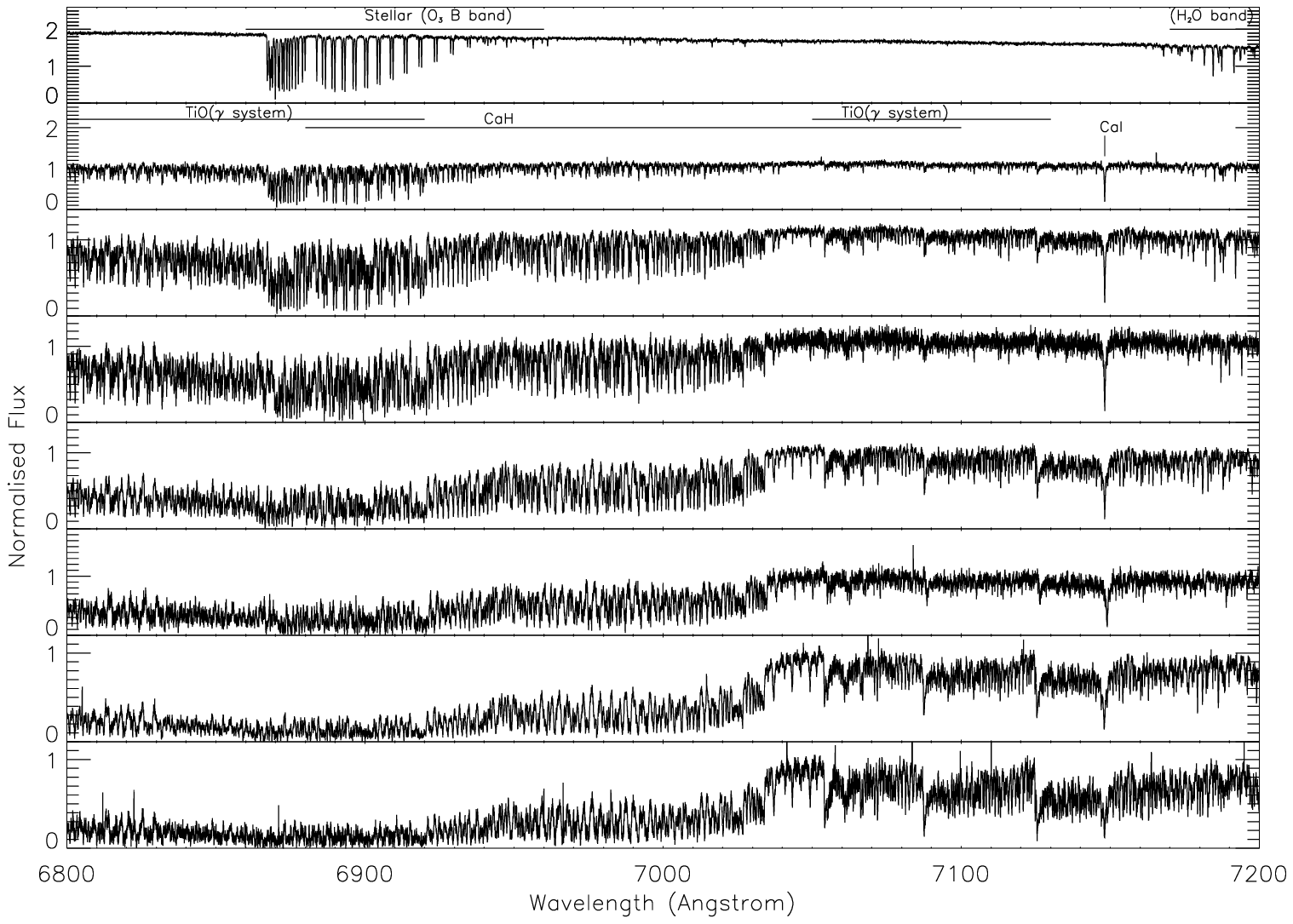}}
\caption{Same as Fig. \ref{Fig:1} for esdM and usdM stars.}
\label{Fig:2}
\end{figure*}

\begin{figure*}[ht!]
\ContinuedFloat
\centering
\subfloat{\includegraphics[width=13.5cm,height=10.0cm]{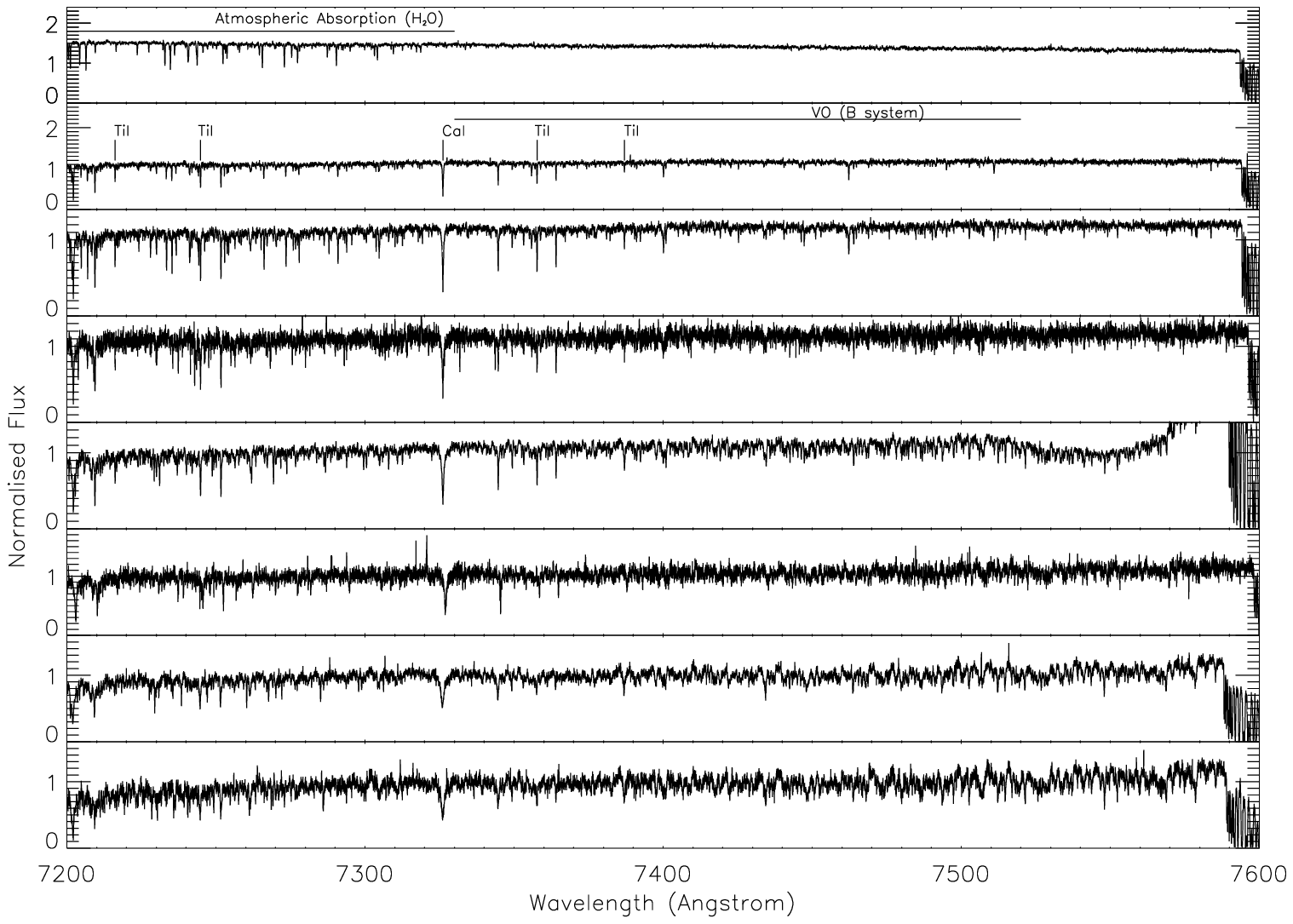}}
\qquad
\subfloat{\includegraphics[width=13.5cm,height=10.0cm]{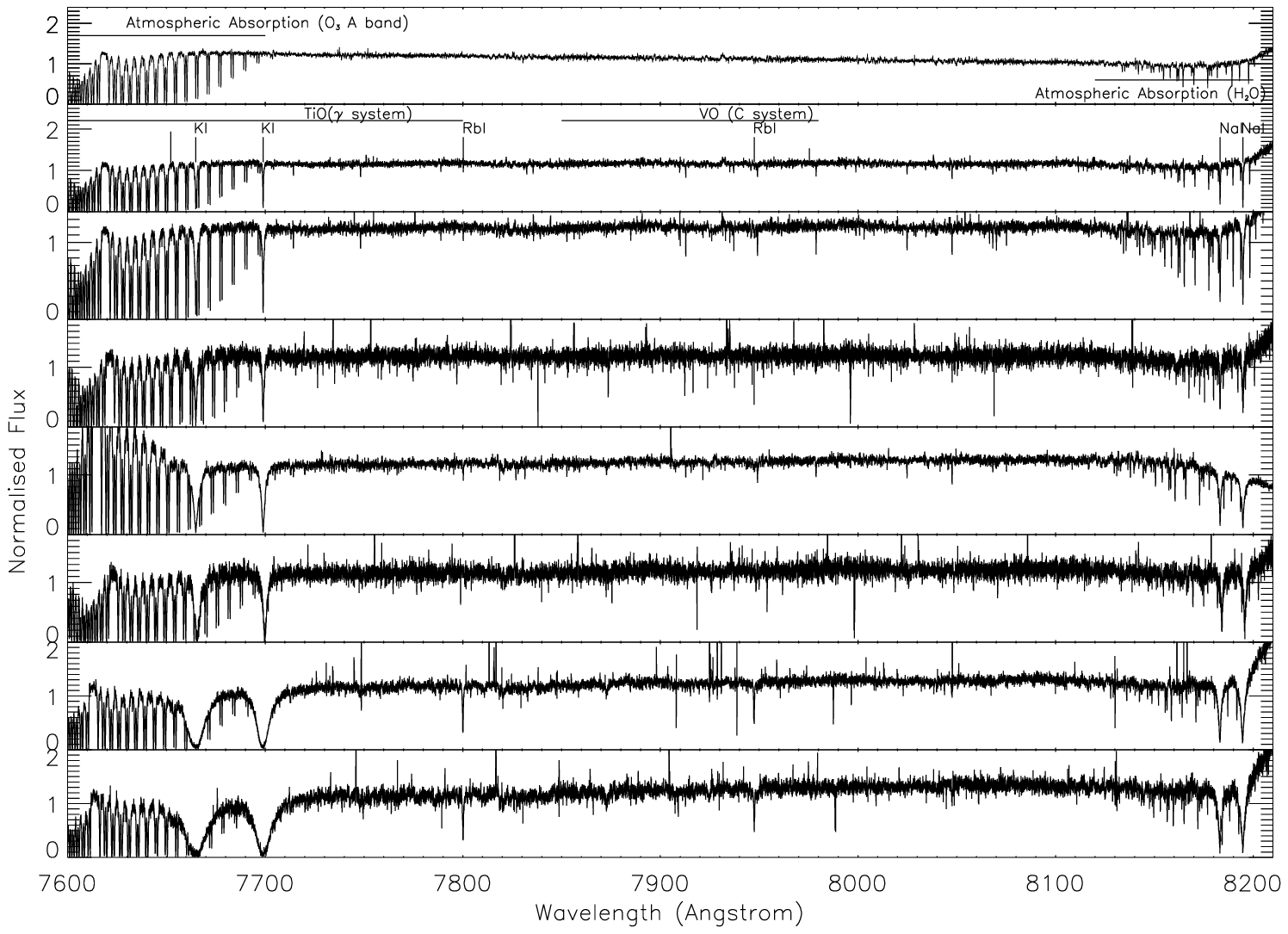}}
\caption{Continued.}
\label{Fig:2}
\end{figure*}

\begin{figure*}[ht!]
\ContinuedFloat
\centering
\subfloat{\includegraphics[width=13.5cm,height=10.0cm]{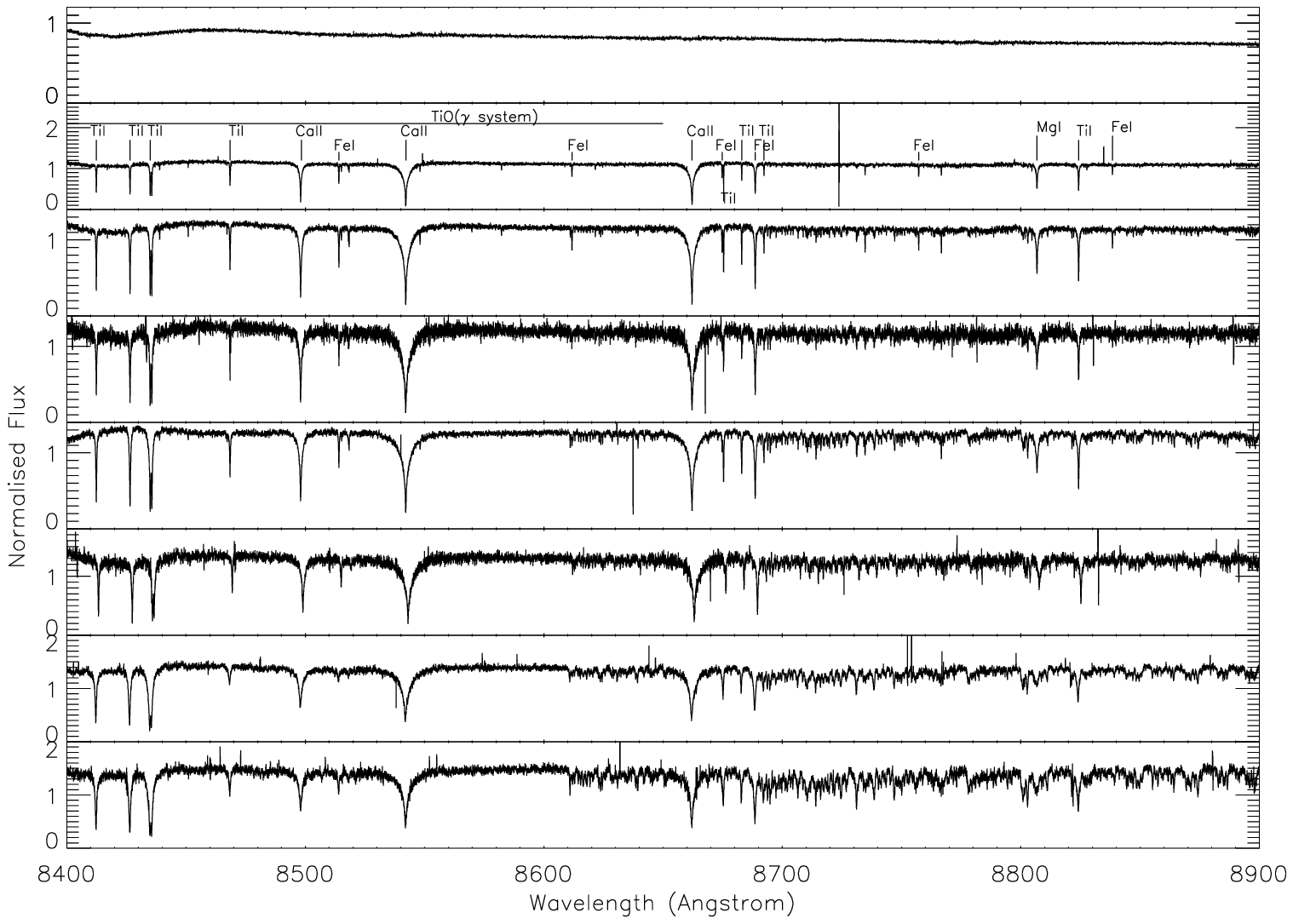}}
\caption{Continued.}
\label{Fig:2}
\end{figure*}

\subsection{Molecular features}
The optical spectra are dominated by molecular absorption bands from metal oxide species such as titanium oxide (TiO), vanadium oxide (VO), and hydrides such as CaH and H$_2$O. They are the most significant opacity sources. However, due to the low metallicity of subdwarfs, they are TiO depleted. The primary effects are the strengthening of hydride bands and collision-induced absorption (CIA) by H$_2$, and the broadening of atomic lines \citep{Allard1997}. Unlike TiO, which produces distinctive band heads degraded on the red, VO produces more diffuse absorption. CaH hydride bands are significant opacity sources, but decrease in relative strength and become saturated with decreasing temperature. 

\subsection{Atomic lines}
All the observed M subdwarfs show strong alkali lines in the observed wavelength range. They are massively pressure broadened, as expected from their high-gravity surface. Atomic features such as Ca II, K I, Rb I, Na I, Ti I, and Mg I are visible throughout the sequence, and their lines are prominent in almost all of the spectra. However, in regions where strong atmospheric absorption is present, it is difficult to measure the intensities of these lines. 

The Na I lines at 8183 $\AA$ and 8194 $\AA$ are clearly visible in all the observed spectra and become broadened from hotter to cooler M subdwarfs. In our setting they appear just at the red end of the lower chip of the blue arm. The K I lines are very narrow for early-type subdwarfs but become very wide and smooth in late-type subdwarfs. The K I resonance lines at 7665 $\AA$ and 7698 $\AA$ govern the spectral shape of cool subdwarf spectra. The equivalent width of these K I lines are of several hundred $\AA$. The ionized Ca II triplet lines at 8498 $\AA$, 8542 $\AA$, and 8662 $\AA$ are very strong in all the observed spectra. Their detailed study by \cite{Mallik1997} shows that their strengths depend on stellar parameters like luminosity, temperature, and metallicity. They are ideal candidates to study their sensitivity to various stellar parameters in cool stars. Although the lower levels of the Ca II triplet lines are populated radiatively and are not collision controlled, they have been identified as very good luminosity probes by \cite{Mallik1997}.  They are relatively free from blends and are little contaminated by telluric lines. In contrast, the Na I lines at 8183 $\AA$ and 8195 $\AA$, which have also been used as luminosity probes in cool stars, have several atmospheric absorption lines in their vicinity \citep{Alloin1989,Zhou1991}. 

\section{Model atmospheres}
\label{S_mod}

%Cool subdwarfs \sout{usually} have low metallicity, and therefore their opacities are different from those of dwarfs. With decreasing temperature, sdM spectra show an increase in abundances of diatomic and triatomic molecules in the optical and in the Near-Infrared (such as SiH, CaH, CaOH, TiO, VO, CrH, FeH, OH, H2O, CO). Oxides such as TiO and VO and hydrides such as CaH, FeH, MgH dominates the opacity sources in the optical and the H$_2$O bands in the Infrared. 

We used the most recent BT-Settl models that were partially published in a review by \cite{Allard2012}. These atmosphere models are computed with
the \texttt{PHOENIX} multi-purpose atmosphere code version 15.5 \citep{Hauschildt1997,Allard2001}, solving the radiative transfer in 1D spherical
symmetry, with the classical assumption of: hydrostatic equilibrium, convection using the mixing-length theory, chemical equilibrium, and a sampling
treatment of the opacities. The models use a mixing length as derived by the radiation hydrodynamic 
simulations of \cite{Ludwig2002,Ludwig2006}, and \cite{Freytag2012}.
%and a radius as determined by the \citep{Baraffe1998} interior models as a function of the atmospheric parameters ($\teff$, $\mathrm{log}\,g$,
%[Fe/H]).

Compared with previous models by \cite{Allard2001}, the current version of the BT-Settl model atmosphere uses the BT2 water-vapor line list computed by \cite{Barber2006}, TiO, VO, CaH line lists by \cite{Plez1998}, MgH by \cite{Weck2003} and \cite{Story2003}, FeH and CrH by \cite{Dulick2003} and \cite{Chowdhury2006}, CO$_2$ by \cite{Tashkun2004}, H$_2$ CIA by \cite{Borysow2001,Abel2011}, and CO by \cite{Goorvitch94a,Goorvitch94b}, to mention the most important line lists. We also included the H$_2$-He CIA from \cite{Borysow1989} and \cite{Borysow1997}, H$_2$-H from \cite{Gustafsson2003}, and H-He (at least in the most recent models) from \cite{Gustafsson2001}. Detailed profiles for the alkali lines were also used \citep{AllardN2007}. The reference solar elemental abundances used in this version of the BT-Settl models those measured by \cite{Caffau2011}.

In general, the \cite{Unsold1968} approximation is used, with the general exception of Na I, Si I, Ca I, and Fe I, where we instead used the respective correction factors found by \cite{Gustafsson2008} for the atomic damping constants with a correction factor to the widths of 2.5 for the non-hydrogenic atoms \citep{Valenti1996}. More accurate broadening data for neutral hydrogen collisions by \cite{Barklem2000} were included for several important atomic transitions such as the alkali, Ca\,I, and Ca\,II resonance lines. For molecular lines, we adopted average values (e.\,g.\ $\langle\gamma_6^{HIT}(T_0, P_0\footnote{Standard temperature  296\,K and pressure 1 atm})\rangle_{H_2O} = 0.08 \,\, P_{\rm gas} \, [\mathrm{cm}^{-1}\mathrm{atm}^{-1}]$ for water vapor lines) from the HITRAN database \citep{HITRAN2008},
which are scaled to the local gas pressure and temperature 
\begin{equation}
\gamma_6(T) = \langle\gamma_6^{HIT}(T_0,P_0)\rangle \left(\frac{296\,K}{T}\right)^{0.5}\, \left(\frac{P}{1\, {\rm atm}} \right) \enspace,
%\enspace. 
\end{equation} 
with a single temperature exponent of 0.5, to be compared with values ranging mainly from 0.3 to 0.6 for water transitions studied by
\cite{Gamache1996}. The HITRAN database gives widths for broadening in air, but \cite{VSTAR2012} found that these agree in general within 10\,--\,20\% with
those for broadening by a solar-composition hydrogen-helium mixture.

The BT-Settl grid extends from $\teff$ = 300 to 7000 K in steps of 100 K, $\mathrm{log}\,g$ = 2.5 to 5.5 in steps of 0.5, and [M/H]= -2.5 to 0.0 in steps of 0.5, which accounts for alpha-element enrichment.  The alpha enhancement was taken as  [alpha/Fe] = -0.4$\times$[Fe/H] for 0 $\textgreater$ [Fe/H] $\textgreater$ -1 and 0.4 for metallicities $\textless$ -1.0, which is also consistent with the choice of \cite{Gustafsson2008}. These different prescriptions for alpha enhancement are rough estimates for the old thin disc and thick disc, respectively \citep[see e.g.][for a discussion on the abundance trends relative to Fe]{Neves2009}.

%The alpha enhancement is 0.2 {\bf{at}} [M/H] = -0.5, and 0.4 {\bf{at or below}} [M/H] = -1.0. 

We linearly interpolated the grid at every 0.1 dex in $\mathrm{log}\,g$ and metallicity. For more details of the BT-Sett model atmosphere see \cite{Allard2012,Allard2013} and\cite{Rajpurohit2012a}. The synthetic colours and spectra were distributed with a resolution of around $R=100 000$ via the \texttt{PHOENIX} web simulator\footnote{http://phoenix.ens-lyon.fr/simulator}.

\begin{figure*}[ht!]
\centering
\includegraphics[width=13.5cm,height=10.0cm]{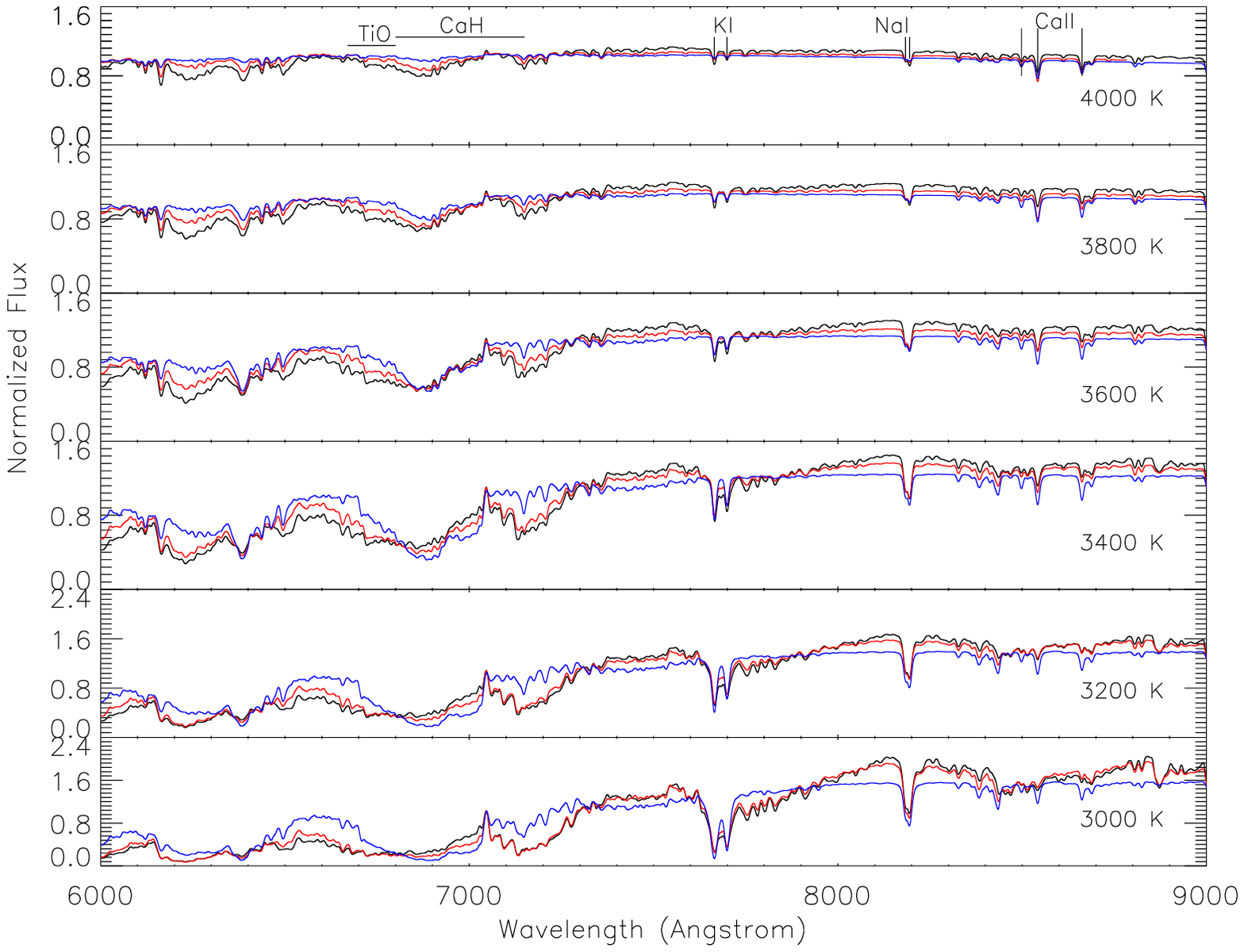}
\caption{ BT-Settl synthetic spectra from 4000 K to 3000 K. The black, red, and blue lines represent [M/H] = 0.0, -1.0, and -2.0, respectively. %The stronger CaH bands at a given temperature could be caused by lower metallicity. This should not be in the caption.
}
\label{Fig:3}
\end{figure*}

Fig. \ref{Fig:3} shows BT-Settl synthetic spectra varying $\teff$ and [M/H].  Oxide bands that dominate in M dwarf spectra are weaker in the subdwarfs where the hydride bands dominate \citep{Jao2008}. They have complex and extensive band structures that leave no window for the true continuum and create a pseudo-continuum that only lets pass the strongest, often resonance atomic lines \citep{Allard1990, Allard1995}. However, because of the lower metallicity of subdwarfs, the TiO bands are weaker, and the pseudo-continuum is brighter. This increases the contrast to the other opacities such as hydride bands and atomic lines, which feel the higher pressures of the deeper layers where they emerge from. We therefore see these molecular bands with more detaile and under more extreme gas pressure conditions than for M dwarfs.

Although the models were computed for different [M/H] values, in the following we use [Fe/H] everywhere. Indeed, we use only the iron lines for the metallicity determinations so that in fact we determine a [Fe/H] value.

\section{Comparison with model atmospheres}
\label{smd}

\begin{figure*}[ht!]
\centering
\includegraphics[width=13.5cm]{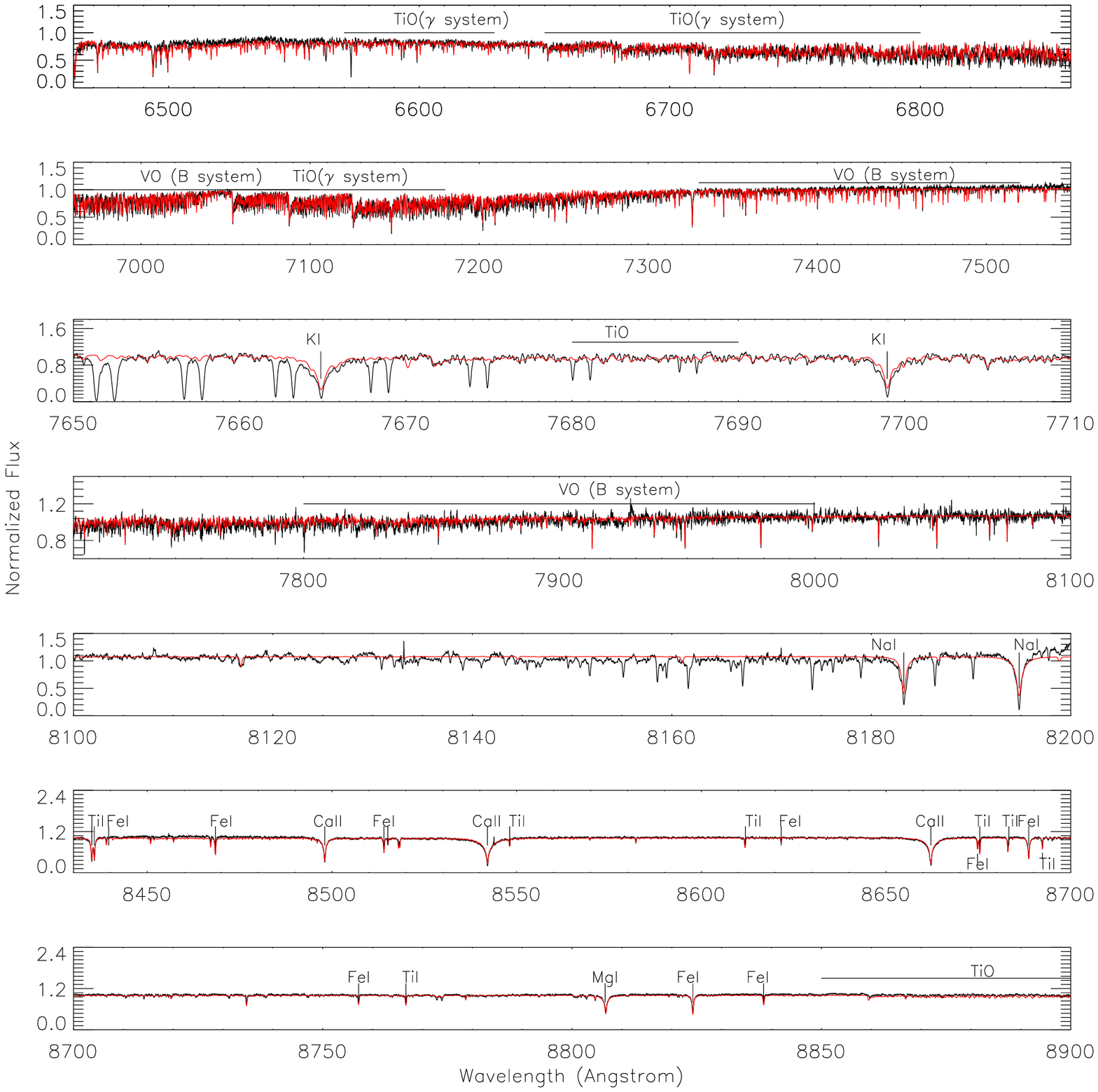}
\caption{UVES spectra of the sdM1 star LHS~158 (black) compared with the best-fit BT-Settl synthetic spectra (red).}
\label{Fig:4}
\end{figure*}

\begin{figure*}[ht!]
\centering
\includegraphics[width=13.5cm]{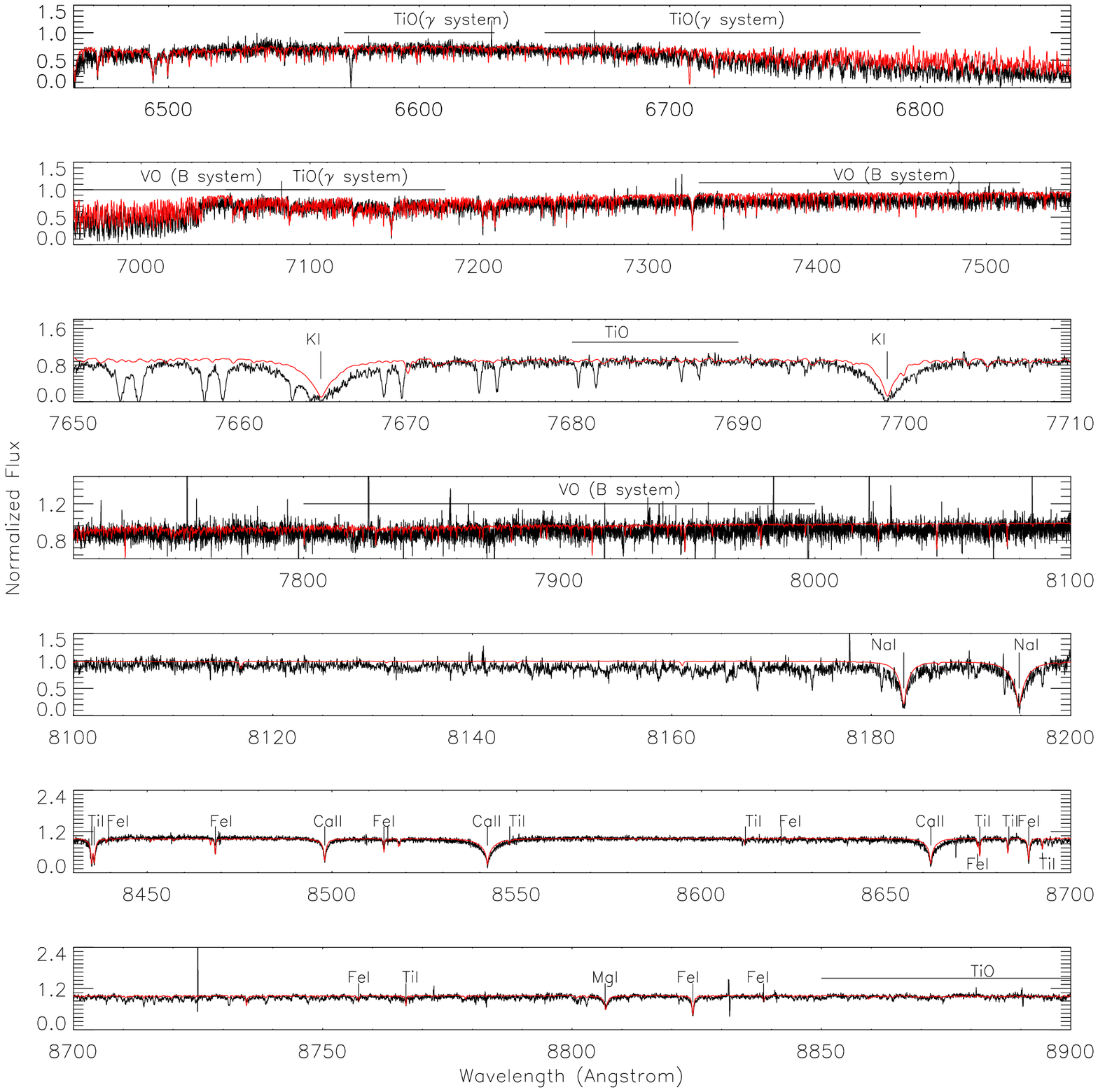}
\caption{Same as Fig.~\ref{Fig:4} for the usdM4.5 star LHS~1032.}
\label{Fig:4b}
\end{figure*}

We perform a comparison between observed and synthetic spectra computed from the BT Settl model to derive the physical parameters of our sample.
Furthermore, the comparison with observed spectra is very crucial to reveal the inaccuracy or incompleteness of the opacities used in the model.  Figures \ref{Fig:4} and \ref{Fig:4b} show the comparison of the best-fit model for an sdM1 and a usdM4.5 star.

\subsection{Molecular bands}
The spectra agree well and reproduce the specific strengths of the TiO band heads at 6600\,$\AA$,  6700\,$\AA$, 7050\,$\AA,  $7680$\,\AA$, and 8859\,$\AA$, and of the VO bands at 7000\,$\AA$, 7430\,$\AA$, and 7852\,$\AA$. The excellent match between models and observations over entire subdwarf sequence shows that the high-frequency pattern visible at this spectral resolution is the structure of the absorption band and not noise.  %However, \cite{Allard1995} shows that  models capture the general overall behavior of metal-poor stars -with decreasing metallicity the hydride bands and atomic lines both increase in strength- the latter due to an increase in gas pressure. 

\subsection{Atomic lines}

The models predict the shape of the Na I doublet at 8194 $\AA$ and 8183 $\AA$ rather well but its strength is well fitted only poorly. In the sdM3.0 and later and in esdM, the observed lines are broader and shallower than those predicted by the models.

%The K I resonance lines at 7665 $\AA$ and 7698 $\AA$ govern the spectral shape of cool dwarf spectra in the optical wavelength regime. These lines have equivalent widths of several hundred angstroms in M dwarfs and are evolving from narrow lines in early sdM and esdM to very wide and smooth absorption \textbf{troughs : what does that mean?}  in late sdM and esdM with wings extending more than 1000 $\AA$ to the blue and especially to the red. 
The qualitative behaviour of the K I doublet at 7665 $\AA$ and 7698 $\AA$ is well reproduced by the models, especially the strong pressure-broadening wings in the early sdM and esdM. 
In the sdM0, the cores of the observed K I lines are still visible as relatively narrow absorption minima embedded in wings extending a few tens to one hundred $\AA$. This broader absorption component becomes saturated in sdM7 spectrum. The models also show this effect but do not reproduce it perfectly. This may be an indication (i) that the models do not yet produce correct densities of the neutral alkali metals in the uppermost part of the atmosphere, since the central parts of the lines form in the highest layers of the atmosphere, especially in the early sdM and esdM where alkali metals are not depleted strongly \citep{Johnas2007}, or (ii) that the depth of atomic lines can only be reproduced when accounting for a magnetic field and Zeeman broadening \citep{Deen2013}.

The models reproduce the strength and wings of Ca II triplet lines at 8498 $\AA$, 8542 $\AA$, and 8662 $\AA$ very well. We also have a good fit  to the Ti I lines. They are located between 8400 $\AA$ to 9700 $\AA$ and belong to low-energy transitions that appear to be visible in such cool atmospheres \citep[see also][] {Reiners2006b}.

Absorption lines of Rb appear in spectra later than sdM7.0. They become stronger and wider towards lower temperature. Given the fact that they are embedded in pseudo-continua that are not always a good fit to the data, the behaviour of the two Rb lines at 7800 $\AA$ and 7948 $\AA$ is very well reproduced by the models.

\subsection{Stellar parameter determination}

The analysis using synthetic spectra requires the specification of several input parameters: effective temperature, surface gravity, and the overall
metallicity with respect to the Sun. We first convolved the synthetic spectrum with a Gaussian kernel at the observed resolution and then interpolated the
result with the observation. We performed a $\chi^2$ fitting between the grid of synthetic spectra and observed spectra in the wavelength range 6400
$\AA$ to  8900 $\AA$, as shown by \cite{Onehag2012}.  We excluded the spectral region between 6860 to 6960 $\AA$, 7550 to 7650 $\AA$, and 8200 to
8430 $\AA$ because of atmospheric absorption. We let all the stellar parameters ($\teff$, [Fe/H], $\mathrm{log}\,g$) vary.

The minimum $\chi^2$ value gives the best-fit parameters and their error bars were derived by allowing a 5 \% deviation of the $\chi^2$ value from its minimum value. The acceptable parameters were finally inspected by comparing them with the observed spectra. 

%The resultant uncertainties is in good agreement with the spectral points used to constrain the fit include the entire TiO band head region and the atomic line profiles. 
%We tested the sensitivity of the TiO lines to surface gravity and did not find any significant effect in the parameter space and wavelength range relevant to this study. 

The TiO system at 6600 $\AA$, 6700 $\AA$, and 7100 $\AA$ is highly sensitive to $\teff$ (increasing in strength with decreasing $\teff$) and rather insensitive to variation in gravity. At a given $\teff$, the band strengths change only slightly even for a large 0.5 change in gravity (in the log g =4.5-5.5 range expected for low-mass stars). At a given gravity, however, they vary significantly over a change of only 100 K in $\teff$. %This TiO system is thus an excellent temperature indicator.

\begin{figure}[ht!]
\centering
\includegraphics[width=8cm]{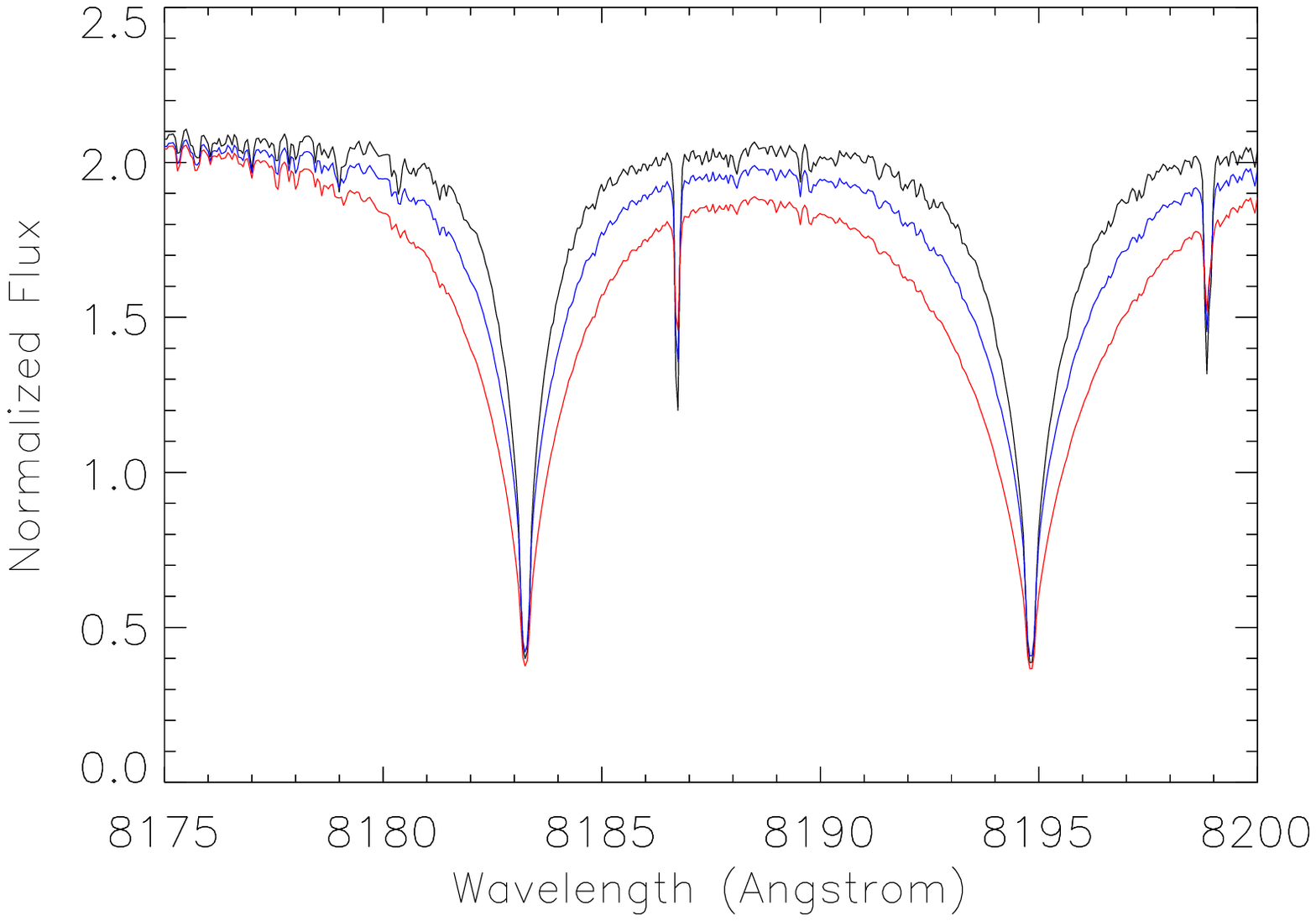}
\caption{BT-Settl synthetic spectra with $\teff$ of 3500 K and varying $\mathrm{log}\,g$ =4.5 (black), 5.0 (blue), 5.5 (red). The effect of gravity and pressure broadening on the sodium doublet is clearly visible.}
\label{Fig:NaI}
\end{figure}

The surface gravity determination was checked on the width of gravity-sensitive atomic lines such as the K I and Na I doublets (see Fig. \ref{Fig:NaI}) as well as on the relative strength of metal-hydride bands such as CaH. The K I doublet at 7665 $\AA$ and 7699 $\AA$  and the Na I lines at 8183 $\AA$ and 8194 $\AA$  are particularly useful gravity discriminants for M dwarfs and subdwarfs. The overall line strength (central depth and equivalent width) increases with gravity as the decreasing ionization ratio due to higher electron pressure leaves more neutral alkali lines in the deeper atmosphere \citep{Reiners2005}. The width of the damping wings in addition increases due to the stronger pressure broadening, mainly by H$_2$, He, and H I collisions. 

%\section{Metallicity}

\begin{figure*}[ht!]
\centering
\includegraphics[width=13.5cm,height=12.0cm]{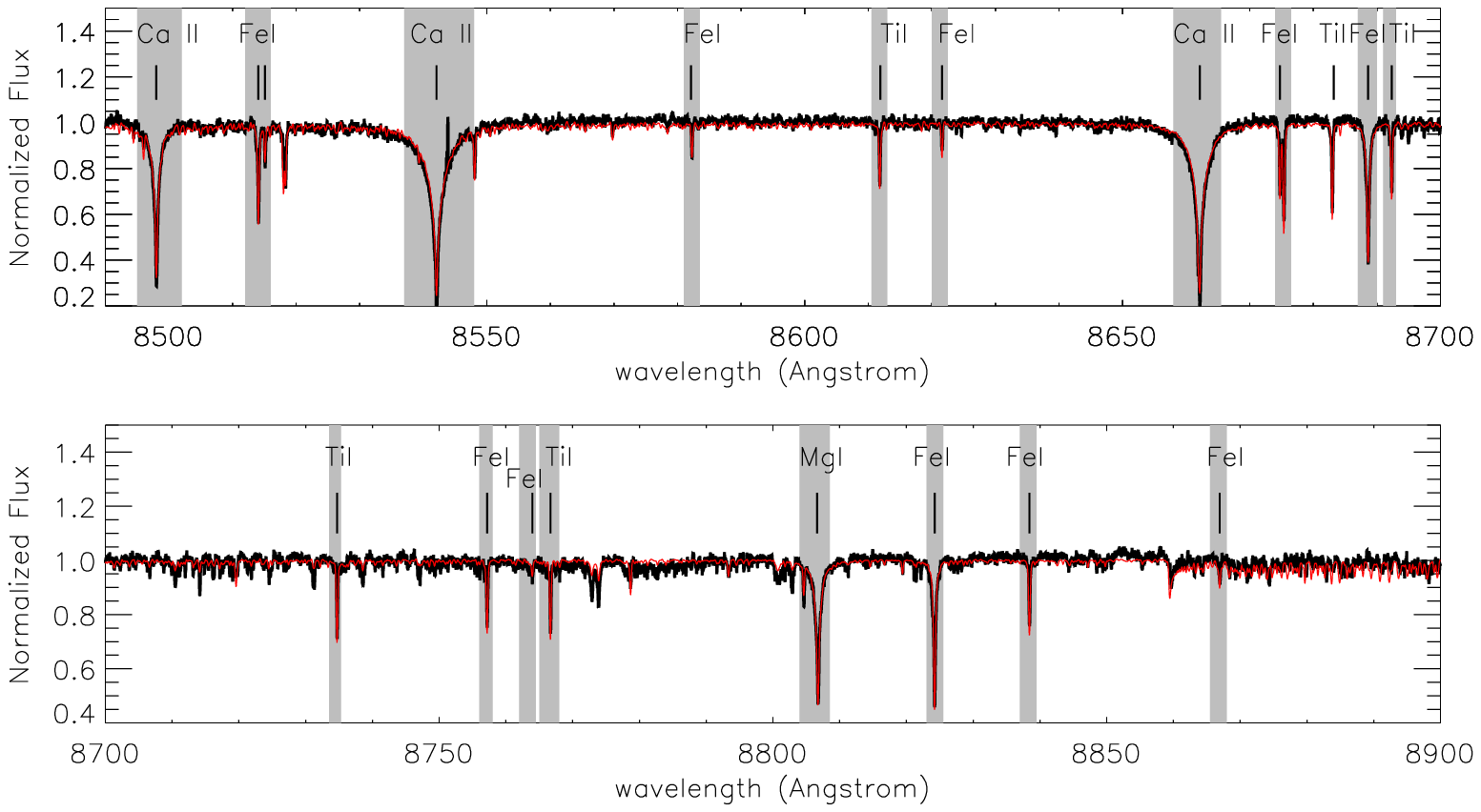}
\caption{UVES spectrum of the sdM1 star LHS 158 (black) and the best-fit BT Settl synthetic spectrum (red). The strong atomic features are highlighted.}
\label{Fig:5}
\end{figure*}

We also checked the metallicity determination on the spectral interval from 8440 $\AA$ to  8900 $\AA$ where molecular absorptions are lower and atomic
lines appear clearly (see Fig. \ref{Fig:5}). The synthetic spectrum represents the line profiles fairly well  for elemental species such
as Ti, I Fe I, Ca II, and Mg I. The best-fit parameters ($\teff$, $\mathrm{log}\,g$, [Fe/H]) are given in Table~2. \\

%For the metallicity determination we fit the synthetic spectra using the same procedure %explained above but on restricted regions where molecular absorptions are less and atomic %lines appear clearly.
%Our first criterion is that the lines must have fractionally small amount of TiO line blending. %This requires the use of fairly strong lines. The second criterion is that they must also be strong %enough in a solar spectrum for the purpose of determining astrophysical values of %$\mathrm{log}\,g$f. Most of the lines we select occupy a short spectral interval from 8440 %$\AA$ to  8900 $\AA$. In our analysis we also include the new lines of elemental species such %as Ca II triplet and Ti lines. In addition to the atomic lines we have also used the TiO bandhead %around 7088 $\AA$ in the analysis of the metallicity determination. 

\begin{table*}
\centering
\caption{Stellar parameters of the observed targets.}
\begin{tabular}{ccccc}
\hline
Target& Spectral Type &   $\teff$ &	  $\mathrm{log}\,g$ &	  [Fe/H]\\
\hline

LHS 72&				sdK4	&	3900	$\pm$23	&4.5$\pm$0.13&	-1.4$\pm$	0.27 \\
LHS 73&				sdK7	&	3800	$\pm$39	&4.5$\pm$0.13&	-1.4$\pm$0.19\\
G 18-37&			esdK7&	3800	$\pm$78	&4.5$\pm$0.15&	-1.3$\pm$0.44\\%SDSS 221455
APMPM J2126-4454&		sdM0&	3700	$\pm$49	&4.5	$\pm$0.19&	-1.3$\pm$0.23\\
LHS 300&				sdM0&	3800	$\pm$39	&4.5$\pm$0.17&	-1.4$\pm$0.24\\
LHS 401&				sdM0.5&	3800	$\pm$26	&4.5$\pm$0.17&	-1.4$\pm$0.28\\
LHS 158&				sdM1&	3600	$\pm$48	&4.5$\pm$0.17&	-1.0$\pm$0.3\\
LHS 320&				sdM2&	3600	$\pm$59	&4.6$\pm$0.23&	-0.6$\pm$0.31\\
LHS 406&				sdM2&	3600	$\pm$40	&4.7$\pm$0.24&	-0.6$\pm$0.24\\
LHS 161&				esdM2$^a$&	3700	$\pm$77	&4.8$\pm$0.30&	-1.2$\pm$0.36\\ %sdK6.0
LP 771-87&		usdM2&	3600	$\pm$95	&4.8$\pm$0.31&	-1.4$\pm$0.52\\%2MASS 030734
LHS 541&			sdM3&	3500	$\pm$76	&5.1$\pm$0.31&	-1.0$\pm$0.39\\
LHS 272&				sdM3&	3500	$\pm$66	&5.2$\pm$0.30&	-0.7$\pm$0.37\\
LP 707-15&			esdM3&	3500	$\pm$68	&5.5$\pm$0.29&	-0.5$\pm$0.36\\%SDSS 010954
LSR J1755+1648&			sdM3.5&	3400	$\pm$52	&5.1$\pm$0.31&	-0.5$\pm$0.33	\\			
LHS 375&				esdM4&	3500	$\pm$79	&5.5$\pm$0.32&	-1.1$\pm$0.31\\
LHS 1032&			usdM4.5&	3300	$\pm$63	&4.5$\pm$0.32&	-1.7$\pm$0.25\\
SSSPM J0500-5406&		esdM6.5&	3200$\pm$51	&5.4$\pm$0.31&	-1.6$\pm$0.16\\
LHS 377&				sdM7&	3100	$\pm$32	&5.3$\pm$0.25&	-1.0$\pm$0.16\\
APMPM J0559-2903&		esdM7&	3200	$\pm$68	&5.4$\pm$0.34&	-1.7$\pm$0.25\\
SSPM J1013-1356&			sdM9.5&	3000	$\pm$0		&5.5$\pm$0.05&	-1.1$\pm$0.16\\

\hline
\end{tabular}

$^a$  The spectral type derived from \cite{Gizis1997} is probably too late because of an observational problem that affects the extreme red slope of the spectrum (J. Gizis, private communication). This agrees with the warmer temperature we derived for this object.
\end{table*}

%
%\begin{figure*}[ht!]
%\centering
%\includegraphics[width=13.5cm,height=12.0cm]{LHS158.ps}
%\caption{Continued.}
%\label{Fig:5}
%\end{figure*}
%
%

\section{Discussion}
\label{disc}

\begin{figure*}[ht!]
\centering
\includegraphics[width=8.0cm]{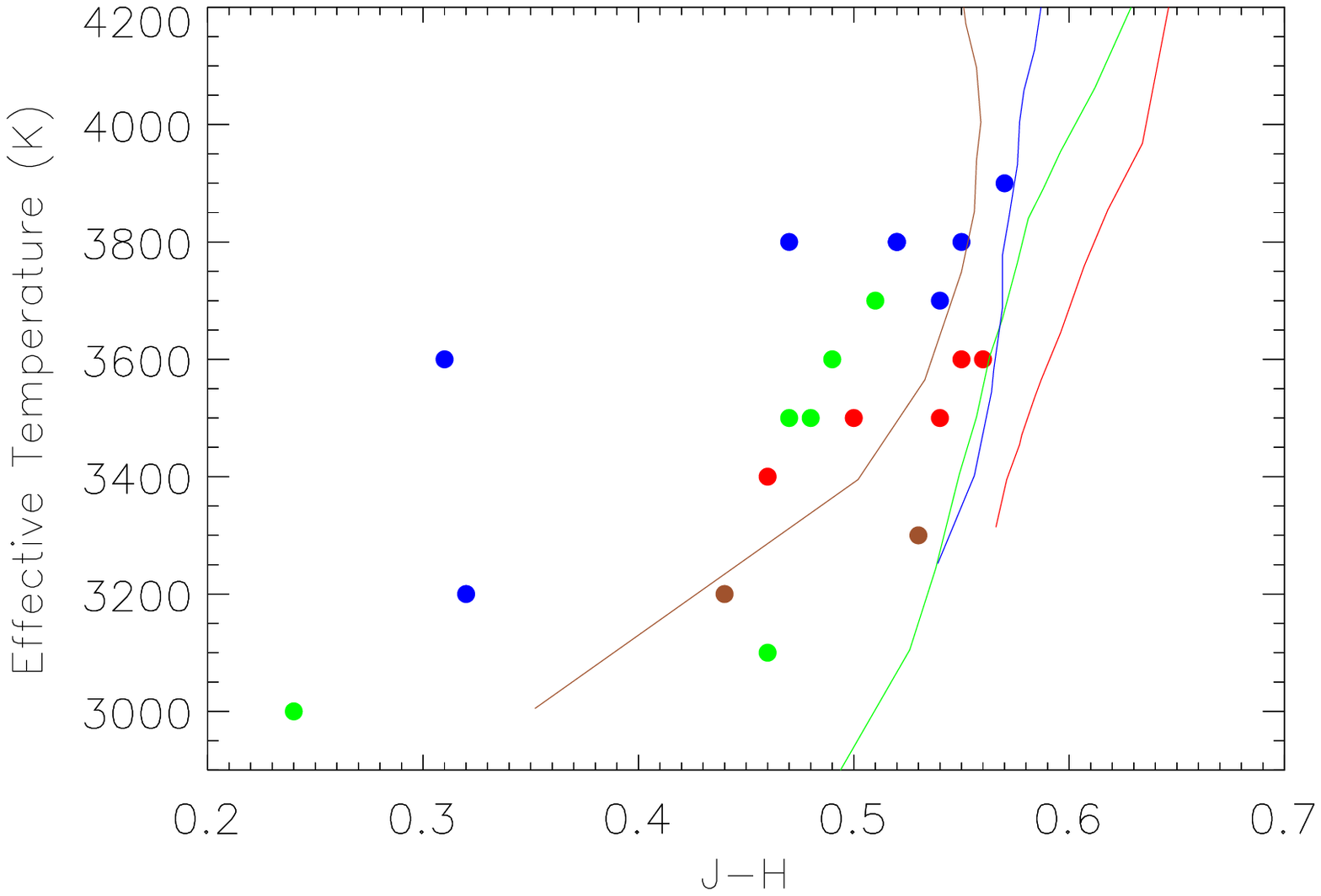}
\includegraphics[width=8.0cm]{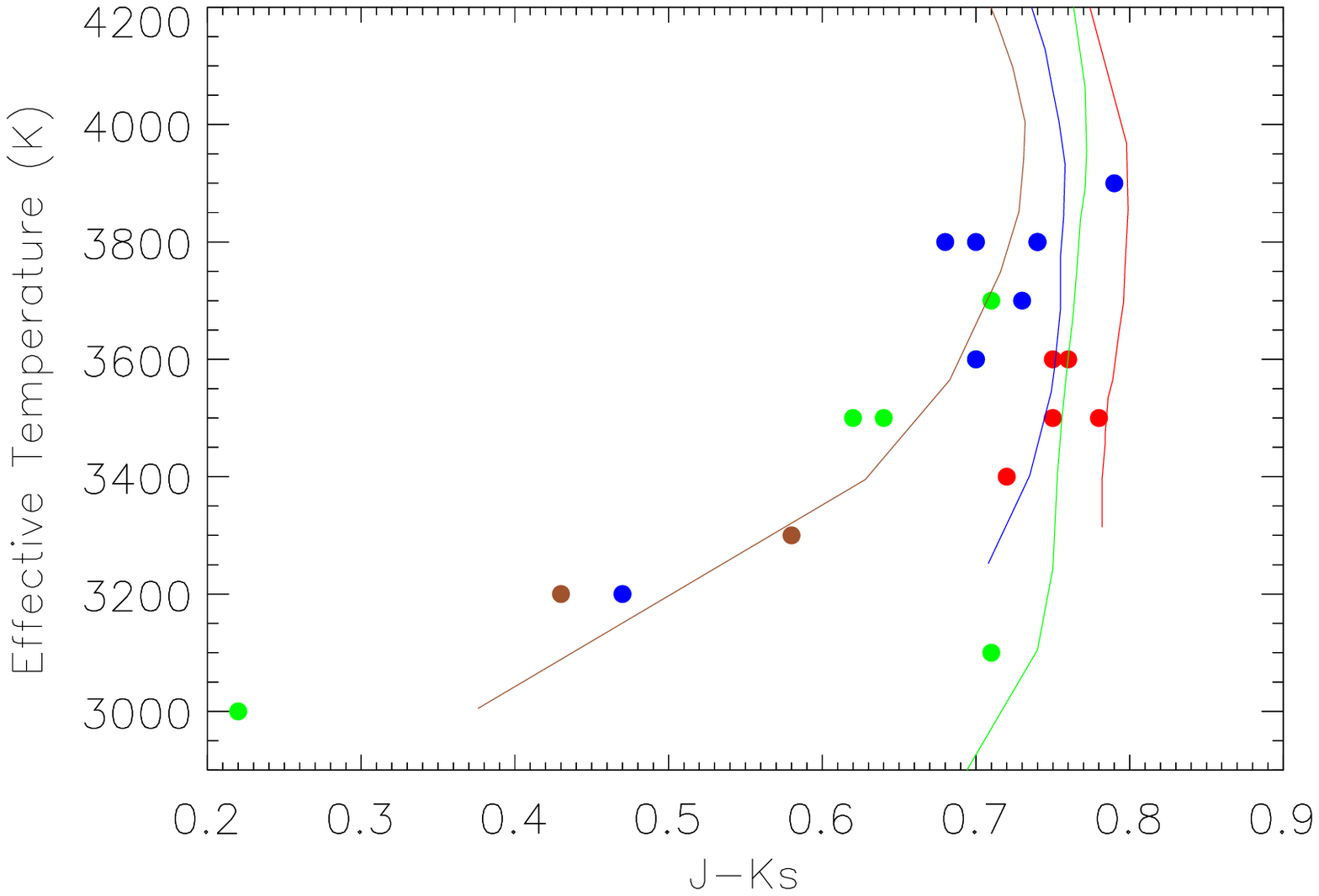}
\caption{Effective temperature versus near-infrared colours of our sample of subdwarfs. The different colours stand for different metallicities: [Fe/H] $= -0.5$ to $-0.7$ dex (red), [Fe/H] $= -1.0$ to $-1.2$ dex (green), [Fe/H] $= -1.3$ to $-1.6$ dex (blue), [Fe/H] $= -1.7$ dex (brown). The lines are from evolution models from \cite{Baraffe1998} at different metallicities (red: $-0.5$ dex, green: $-1.0$ dex, blue: $-1.5$ dex, brown: $-2.0$ dex) assuming an age of 10 Gyr. }
\label{Fig:teffcol}
\end{figure*}

The effective temperature versus near-infrared colours are shown in Fig. \ref{Fig:teffcol}. The expected relations from evolution models
\citep{Baraffe1997,Baraffe1998} assuming an age of 10 Gyr and varying metallicities are also superimposed. The colours stand for different
metallicities. The plot shows discrepancies between the models and our observations. It shows that the metallicities determined from the high-resolution spectral features are inconsistent with those we would infer from near-infrared colours. This inconsistency may be due to uncertainties in
the CIA opacities or to outdated model interiors. It shows that broad-band colours are not sufficient to determine the parameters of M subdwarfs.

\begin{figure}[ht!]
\centering
\includegraphics[width=8.0cm]{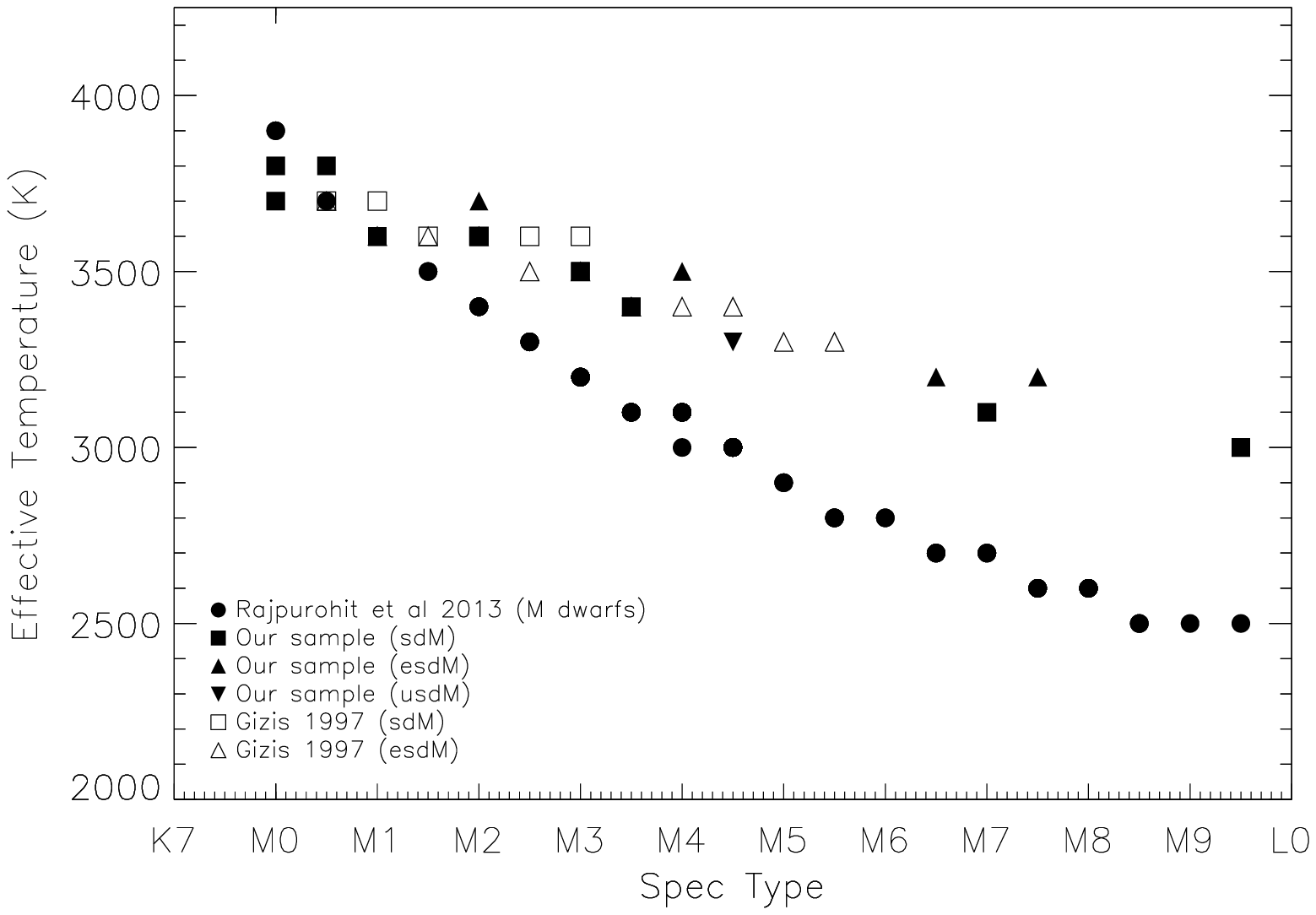}
\caption{Relation of the effective temperature of subdwarfs versus spectral type from our sample (filled symbols) compared with those from \cite{Gizis1997} (open symbols) and the M dwarf $\teff$ scale from \cite{Rajpurohit2013} (filled circles). %and young M dwarfs $\teff$ scale from \cite{Bayo2011} (open circles).
}
\label{Fig:teff}
\end{figure}

The relation of effective temperature versus spectral-type is shown in Fig. \ref{Fig:teff}. The relation determined using UVES sample is compared with the $\teff$ scale of M dwarfs determined by \cite{Rajpurohit2013}. The $\teff$ of subdwarfs is 200-300 K higher than $\teff$ of M dwarfs for the same spectral type except for hot temperatures. This is expected since the TiO bands are depleted with decreasing metallicity, and as a result, the pseudo-continuum is brighter and the flux is emitted from the hot deeper layer. A comparison with the earlier work from \cite{Gizis1997} is also shown. \cite{Gizis1997} determined the temperature by comparing the low-resolution optical spectra of a sample of sdM and esdM with the NextGen model atmosphere grid by \cite{Allard1997}. The $\teff$ for subdwarfs agrees within 100 K. This difference is due to the incompleteness of the TiO and water vapor line lists used in the NextGen model atmospheres compared with the new BT-Settl models. Furthermore, this work allows us to extend the relation to the coolest M subdwarfs. 
%We also compare the  $\teff$ of very young M dwarfs (5-10 Myrs old) obtained from low resolution optical spectra by \cite{Bayo2011}. We found they occupy the similar space as that of subdwarfs, as expected for young M dwarfs that still undergo heating from their gravitational contraction. 
  
We compiled from the literature the spectral indices of TiO5, CaH2, and CaH3 computed from TiO and CaH band strengths on low-resolution spectra  (see Table~1). The last column gives the references for spectral types and spectral indices. When possible, we took both the spectral type and the indices from the same paper. 
We note that the spectral indices may vary over several subtypes for some of the stars from one author to the other. Since \cite{Jao2008} did not give new
spectral types in the scheme developed by \cite{Gizis1997} that was extended by \cite{Lepine2007} (sd, esd, usd), which we used, we adopted the indices from \cite{Jao2008} and assigned sdM types only when no indices were available from the papers providing an sd, esd, usd classification.

\begin{figure}[ht!]
\centering
\includegraphics[width=12.0cm,bb=40 50 650 450,clip=true]{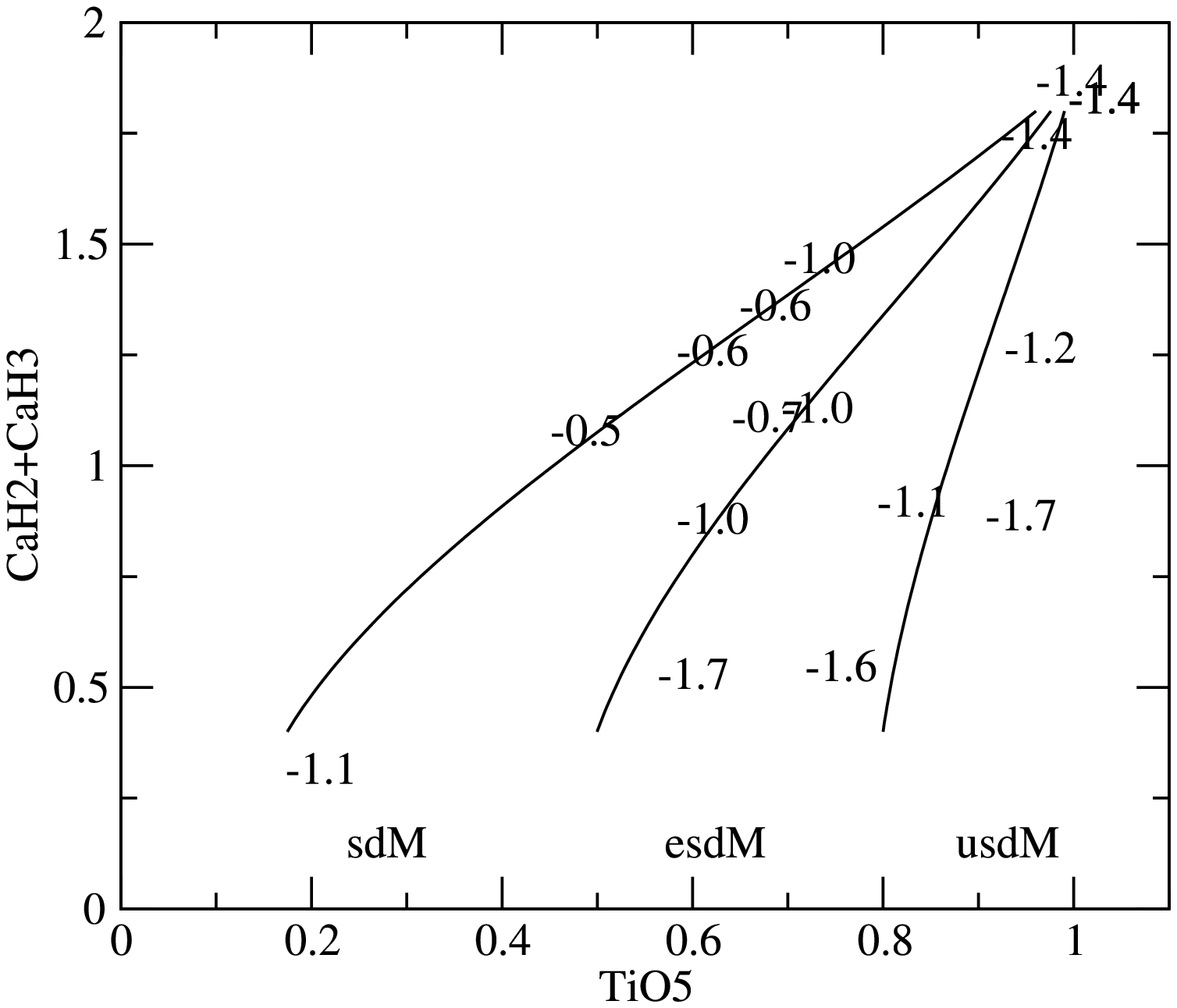}
\caption{CaH2+CaH3 versus TiO5 diagram for our sample. The labels indicate our metallicity determination. The lines are defined by \cite{Lepine2007}. They show the different regions in the diagram where sdM, esdM, and usdM stars are expected to be.}
\label{Fig:class}
\end{figure}

Fig.~\ref{Fig:class} shows the CaH2+CaH3 versus TiO5. \cite{Lepine2003} showed that such a diagram is useful to distinguish between the different object classes, sdM, esdM, and usdM. Our metallicity determinations are labelled in the diagram. It shows that the metallicity decreases from sdM to usdM as expected.

\begin{figure}[ht!]
\centering
\includegraphics[width=8.0cm]{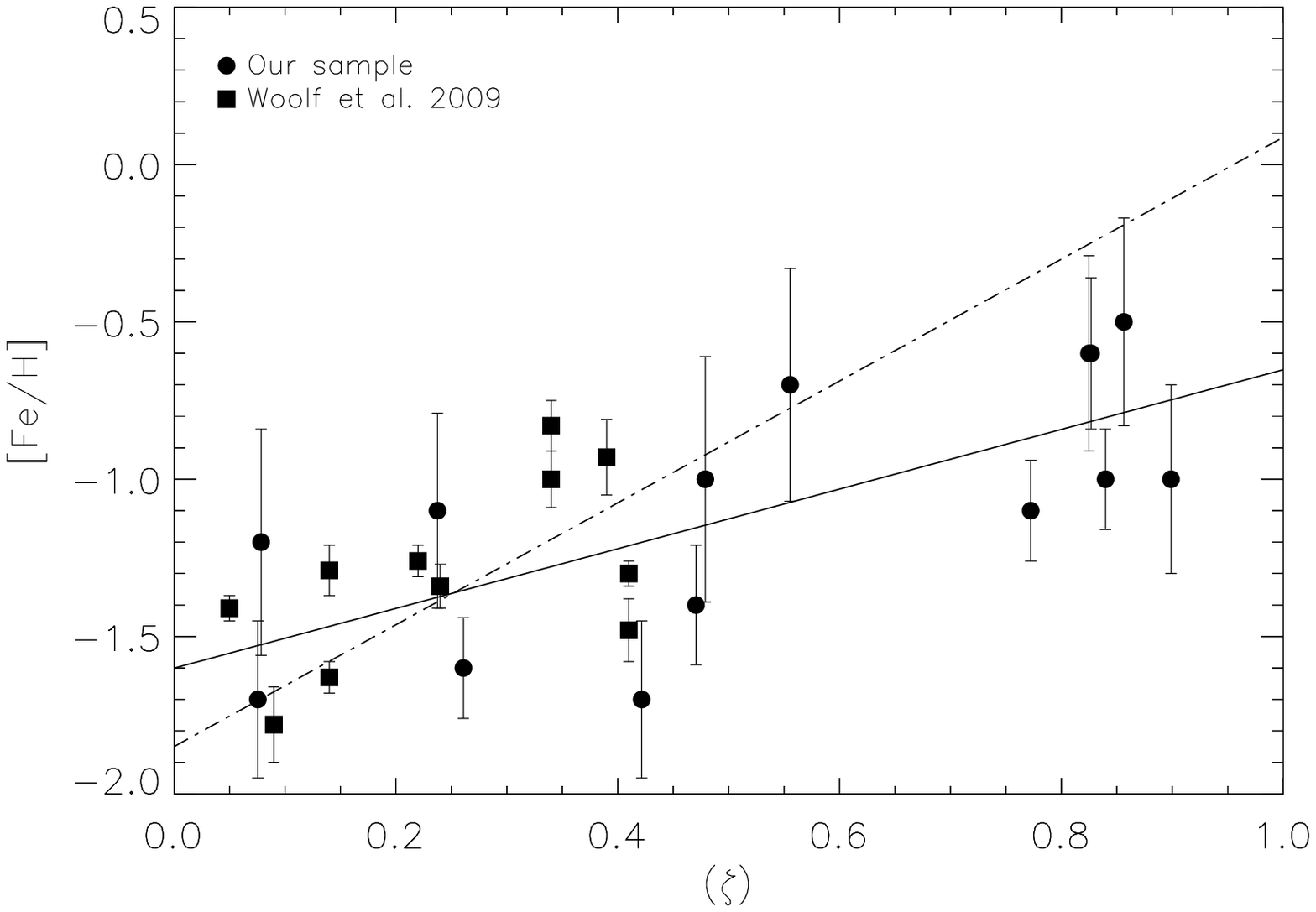}
\caption{$\zeta$ parameter defined by \cite{Lepine2007} versus metallicity diagram. The circles show our sample and the solid black line is the linear-square regression. The sample from \cite{Woolf2009} is superimposed (squares) as well as their derived relation (dashed line).}
\label{Fig:zeta}
\end{figure}

We also derived the $\zeta$ parameter as defined by \cite{Lepine2007}. This $\zeta$ parameter is a combination of the TiO$_5$, CaH$_2$, and CaH$_3$ spectral indices. \cite{Woolf2009} used a sample of 12 esdM and usdM with known metallicities to derive a correlation between $\zeta$ and the metallicity, and showed that it can be used as a metallicity indicator. %\cite{Lepine2013} also gave a correlation on a sample of M dwarfs having metallicity determinations from \cite{Neves2012,Roja2012}. 
We plot the metallicity of our stars versus $\zeta$ (Fig. \ref{Fig:zeta}) and superimpose the sample from \cite{Woolf2009} as well as their derived correlation (dashed line). We also define a similar correlation (solid line) using our sample, $$[Fe/H] = 1.19 \zeta - 1.75$$. 

Although the relation shows a strong dispersion in both samples, our metallicity determinations tend to be lower on average. This can be due to the different spectral indices adopted, or to the use of different atmosphere models. 
LHS 161, the only common star in both samples, illustrates the first assumption. Our metallicity determination (-1.2 dex) agrees with that of \cite{Woolf2009}, (-1.3 dex) but the spectral indices adopted are very different: we used $\zeta=0.07$ from the indices of \cite{Gizis1997} whereas \cite{Woolf2009} used $\zeta=0.41$ from the indices of \cite{Jao2008}. Furthermore, \cite{Woolf2009} used NextGen models.
%that do not account for any alpha enhancement and are based on very different solar abundances from \cite{Grevesse1993} \textbf{Derek, France, more
%to tell?}.
 
%Contrarily to our correlation computed only on a M subdwarfs sample, 
%\cite{Woolf2009} sample contains many M dwarfs which deviate the relation towards the higher metallicity.
%Therefore, this work can contribute to refine the relation for subdwarfs. 
%The solid line shows the linear least-square fit to our sample: $$[Fe/H] = -1.48 \zeta + 0.83$$
%The fit is [Fe/H] = a+b($\zeta$), where a = -1.4759$\pm$0.1979 and b=0.827471$\pm$0.3289.

\section{Conclusion}
\label{ccl}

We presented a high-resolution optical spectral atlas for a sample of 18 M subdwarfs and 3 K subdwarfs, including 5 esdK,1 esdM and 2 usdM\footnote{The
atlas will be made  public through the Virtual Observatory tools.}.  We described various atomic and molecular features that appear in the spectra and
their evolution with decreasing effective temperature and metallicity. We used the most recent BT-Settl model atmospheres, with revised solar
abundances, to determine the scaled solar abundances of the subdwarfs. We compared them with the synthetic spectra produced by the BT-Settl 
atmosphere models and derived their fundamental stellar parameters. The accuracy of the atmospheric models involved in the metallicity determination can be
inferred by looking at the fit to the individual atomic and molecular lines. These high-resolution spectra allowed us to
separate the atmospheric parameters (effective temperature, gravity, metallicity), which is not possible when using broad-band photometry.

We determined the relation of effective temperature versus spectral type of M-subdwarfs and compared it with the previous study from \cite{Gizis1997}. Our relation agrees within 100 K and extends to the cooler spectral sequence. This work also contributes to calibrating the relation between metallicity and photometric colours and molecular band strengths. With calibration, it will be possible to estimate the metallicity of a large sample of subdwarfs, and obtain a meaningful statistical analysis.

The next step of this work is to investigate the effect of different alpha enhancements on the spectral energy distribution and the spectral-line strengths of M subdwarfs. This effect is expected to be as important as the revised oxygen abundances were to make the models fit solar metallicity M dwarfs. This is a  limitation of the models because M subdwarfs are divided into two regimes, each with a rough prescription of alpha enhancements. More realistic abundance trends need to be applied to the models \citep{Neves2009,Adibekyan2012}. This atlas is appropriate for testing these prescriptions and their effect on the atmosphere spectra because they are precise enough to allow us to derive elemental abundances.

\begin{acknowledgements}
We acknowledge observing support from the ESO staff. We acknowledge financial support from "Programme National de Physique Stellaire" (PNPS) of CNRS/INSU, France. We thank the referee J. Gizis for his useful comments on the paper.
\end{acknowledgements}

\bibliographystyle{aa}
\bibliography{ref}
\end{document}